\documentclass[iop]{emulateapj}

\usepackage{natbib,aas_macros,amsmath}
\usepackage{multirow,color}
\usepackage{natbib}

\newcommand{\carcsec}{$\!\!\arcsec$}
\newcommand{\m}[1]{\mathrm{#1}}
\newcommand{\bluec}[1]{\textcolor{black}{#1}}
\newcommand{\redc}[1]{\textcolor{black}{#1}}

\begin{document}
\shortauthors{Harikane et al.}
\slugcomment{Accepted for Publication in ApJ}

\shorttitle{
Absorption Line Spectra of $z\sim6$ LBGs
}

\title{
The Mean Absorption Line Spectra of a Selection of \\Luminous $z\sim6$ Lyman Break Galaxies
}

\email{y.harikane@ucl.ac.uk}
\author{
Yuichi Harikane\altaffilmark{1,2,3},
Nicolas Laporte\altaffilmark{4,5},
Richard S. Ellis\altaffilmark{1},
and Yoshiki Matsuoka\altaffilmark{6}
}

\affil{$^1$
Department of Physics and Astronomy, University College London, Gower Street, London WC1E 6BT, UK
}
\affil{$^2$
National Astronomical Observatory of Japan, 2-21-1 Osawa, Mitaka, Tokyo 181-8588, Japan
}
\affil{$^3$
Institute for Cosmic Ray Research, The University of Tokyo, 5-1-5 Kashiwanoha, Kashiwa, Chiba 277-8582, Japan
}
\affil{$^4$
Kavli Institute for Cosmology, University of Cambridge, Madingley Road, Cambridge CB3 0HA, UK
}
\affil{$^5$
Cavendish Laboratory, University of Cambridge, 19 JJ Thomson Avenue, Cambridge CB3 0HE, UK
}
\affil{$^6$
Research Center for Space and Cosmic Evolution, Ehime University, Bunkyo-cho, Matsuyama, Ehime 790-8577, Japan
}

\begin{abstract}
We examine the absorption line spectra of a sample of 31 luminous ($M_\mathrm{UV} \simeq -23$) Lyman break galaxies at redshift $z\simeq6$ using data taken with the FOCAS and OSIRIS spectrographs on the Subaru and GTC telescopes. For two of these sources we present longer exposure data taken at higher spectral resolution from ESO's X-shooter spectrograph. Using these data, we demonstrate the practicality of stacking our lower resolution data to measure the depth of various interstellar and stellar absorption lines to probe the covering fraction of low ionization gas and the gas-phase and stellar metallicities near the end of the era of cosmic reionization. From maximum absorption line depths of Si{\sc ii}$\lambda$1260 and {\sc Cii}$\lambda$1334, we infer a mean covering fraction of $\ge0.85 \pm 0.16$ for our sample. This is larger than that determined using similar methods for lower luminosity galaxies at slightly lower redshifts, suggesting \redc{that the most luminous galaxies appear to have a lower escape fraction than fainter galaxies, and therefore may not play a prominent role in concluding reionization}. Using various interstellar absorption lines we deduce gas-phase metallicities close to solar indicative of substantial early enrichment. Using selected stellar absorption lines, we model our spectra with a range of metallicities using techniques successfully employed at lower redshift and deduce a stellar metallicity of $0.4^{+0.3}_{-0.1}$ solar, consistent with the stellar mass - stellar metallicity relation recently found at $z\sim3-5$.  We discuss the implications of these metallicity estimates for the typical ages of our luminous galaxies and conclude our results imply initial star formation at redshifts $z\sim10$, consistent with independent analyses of earlier objects.
\end{abstract}


\section{Introduction}\label{ss_intro}

Spectroscopy remains a fundamental tool for understanding the physical processes which govern the evolution of high redshift star-forming galaxies \citep{2016ARA&A..54..761S}. Emission line measurements have been effective not only in supplying galaxy redshifts essential for time-slicing deep survey data, but also in interpreting the nature of their radiation field and analyzing gas-phase metallicities (e.g. \citealt{2006ApJ...644..813E}, \citealt{2016ApJ...831L...9N}, \citealt{2018ApJ...855...42S}, \citealt{2020A&A...636A..47S}). Although more challenging observationally, absorption line spectroscopy can provide additional physical constraints including evidence of kinematic outflows (\citealt{2010ApJ...717..289S}, \citealt{2018ApJ...863..191J}), the chemistry and ionization state of the interstellar gas (\citealt{2012ApJ...751...51J}) and, ultimately, stellar metallicities (\citealt{2019ApJ...885..100L}, \citealt{2019MNRAS.487.2038C}).  

With the exception of studies of rare luminous or gravitationally-lensed examples (e.g. \citealt{2005ApJ...630L.137D}, \citealt{2013ApJ...779...52J}, \citealt{2016ApJ...831..152L}), most progress in absorption line spectroscopy of high redshift galaxies has necessarily involved stacked spectra of representative samples. Following the pioneering study of over 800 Lyman break galaxies (LBGs) at redshift $z\sim3$ by \citet{2003ApJ...588...65S}, \citet{2012ApJ...751...51J} produced a mean rest-frame UV spectrum utilising 80 LBGs at $z\sim4$ and, more recently, \citet{2019MNRAS.487.2038C} analyzed composite spectra drawn from a sample of 681 galaxies in the VANDELS survey spanning a range $2.5<z<5$ with a mean redshift of $\overline{z}=3.5$. Redshift-dependent trends in the stacked spectra of LBGs matched in UV luminosity and stellar mass over $2<z<4$ based on these earlier campaigns are discussed by \citet{2018ApJ...860...75D}. 

One of the most interesting trends, discussed by both \citet{2013ApJ...779...52J} and \citet{2018ApJ...860...75D} is that the equivalent width of low ionization species (LIS) absorption decreases with increasing redshift, possibly due to a reduced covering fraction of neutral hydrogen, $f_\mathrm{c}$, and, by implication an increased escape fraction, $f_\mathrm{esc}$, of ionizing radiation. \bluec{The latter deduction follows \citet{2018A&A...616A..30C} who find a correlation between $f_\mathrm{esc}$ and $1-f_\mathrm{c}$ estimated from LIS absorption using $z\sim0$ Lyman continuum leakers.} \bluec{On the other hand,} \citet{2016ApJ...828..108R} argue that, while the covering fraction of LIS absorption may be a reasonable proxy for that of neutral hydrogen, it may be an unreliable estimate of $f_\mathrm{esc}$ if metal-enriched outflowing gas has a dust content that varies with the covering fraction of hydrogen. Nonetheless, \citet{2018ApJ...860...75D} conclude the evolution of LIS absorption to $z\sim4$ likely represents an increase in the ionizing capability of higher redshift galaxies with important consequences for cosmic reionization. At redshifts $z < 3$ where the escape fraction can be directly measured, the average for both LBGs and metal-poor Lyman alpha emitters (LAEs) is $f_\m{esc}\simeq5-8\%$ (\citealt{2018ApJ...869..123S}, \citealt{2019ApJ...878...87F}). Based on the census of early galaxies, an average $f_\m{esc}>10$\% is necessary to complete cosmic reionization by $z\simeq6$ (\citealt{2013ApJ...768...71R}). An increase in the average $f_\m{esc}$ with redshift would thus be an important result, perhaps indicating earlier galaxies have higher star formation rate surface densities capable of creating porous channels in the interstellar medium (\citealt{2014MNRAS.442.2560W}).

Of course it is unlikely that galaxies of different luminosities and stellar masses have similar covering fractions of hydrogen and LIS gas. In this respect to concept of an {\it average escape fraction} is perhaps naive. Recently there has been much discussion in the literature on the apparent rapidity with which the neutral gas in the intergalactic medium (IGM) became ionized. Estimates of the neutral fraction $x_\m{HI}$, including those from the Gunn-Peterson troughs and proximity zones seen in QSO absorption spectra and the fraction of Lyman $\alpha$ emission seen in color-selected $z>6$ galaxies, collectively imply a rapid evolution from $x_\m{HI}\simeq0.9$ at $z\simeq7.5$ to zero at $z\simeq6$ i.e. within a time interval of only 250 Myr (\citealt{2020ApJ...892..109N}). Such an abrupt ending of reionization may indicate a contribution of ionizing photons from the rarer, more massive, systems which assemble during at the end of the reionization epoch. 

The present paper is motivated, in part, to test this hypothesis by examining the absorption spectra of LBGs at a redshift $z\simeq6$.  \bluec{It is not yet practical with current facilities to study a representative sample of galaxies down to the luminosity limit $M_{UV}\simeq-18$ which Naidu et al propose might be responsible for concluding reionization. However, by analysing the spectra of a more luminous subsample we can at least test whether their escape fractions are large compared to expectations based on reionization being driven uniformly from galaxies across the entire luminosity function}. To date there has been no study at this epoch comparable to those undertaken by the above cited workers at $z\sim2-4$. We present spectra taken with both X-shooter at the ESO Very Large Telescope (VLT) as well as a sample of 31 galaxies taken with Faint Object Camera and Spectrograph \citep[FOCAS;][]{2002PASJ...54..819K} and Optical System for Imaging and low-intermediate-Resolution Integrated Spectroscopy \citep[OSIRIS;][]{2000SPIE.4008..623C} at the Subaru and Gran Telescopio Canarias (GTC) telescopes, respectively. We analyze both individual spectra as well as a composite example to determine the nature of LIS features at the highest redshift for which absorption line work is currently practical.

Such absorption line spectra can also be used to examine the {\it stellar metallicity} of galaxies at high redshift and, particularly, the stellar mass - stellar metallicity relation (\citealt{2019MNRAS.487.2038C}). This is a more attractive, although admittedly more challenging, observational target than that based on gas-phase metallicites, for which there remains debate in the literature about the merits of using local calibrations for the various accessible emission lines, as well as the relative importance of composition and the ionizing radiation field \citep{2014ApJ...795..165S}. Indeed, gas-phase metallicities beyond redshifts $z\sim4$ vary widely according to the technique used. \citet{2017ApJ...846L..30S} estimate the metallicity of a luminous galaxy at $z=4.4$ to be $Z=0.2 Z_{\odot}$ using the [Ne{\sc iii}]/[O{\sc ii}] ratio, suggesting a source in the process of enrichment, whereas \citet{2016ApJ...822...29F} report near solar metallicities at $z\sim5$ based on empirically calibrated relations between absorption line equivalent widths and metallicities. 

A plan of the paper follows. In \S2 we discuss the selection of LBGs which form the basis of the present analysis as well as the spectroscopic observations and associated data reduction. In \S3 we use a cross-correlation technique to determine the individual redshifts prior to producing composite spectra and their uncertainties. In \S4 we discuss our spectra in terms of the covering fraction of LIS absorption and investigate the reliability of using absorption line depths based on our lower resolution Subaru and GTC spectra via comparisons with the X-shooter data. We also derive both stellar and interstellar based metallicities and compare these with mass-dependent relationships available at lower redshifts. In \S5 we discuss briefly the implications of our results both for recent models of late cosmic reionization and constraints on the age and earlier enrichment history of our sample.

Throughout this paper we use the recent Planck cosmological parameter sets constrained with the temperature power spectrum, temperature-polarization cross spectrum, polarization power spectrum, low-$l$ polarization, CMB lensing, and external data \citep[TT, TE, EE+lowP+lensing+ext result; ][]{2016A&A...594A..13P}:
$\Omega_\m{m}=0.3089$, $\Omega_\Lambda=0.6911$, $\Omega_\m{b}=0.049$, and $h=0.6774$.
We assume the \citet{1955ApJ...121..161S} initial mass function (IMF) with lower and upper mass cutoffs of $0.1\ \m{M_\odot}$ and $100\ \m{M_\odot}$, respectively.
All magnitudes are in the AB system \citep{1983ApJ...266..713O}, and are corrected for Galactic extinction \citep{1998ApJ...500..525S}.
The definition of the solar metallicity is given by $12+\m{log(O/H)}=8.69$ and $Z_\odot=0.0142$ \citep{2009ARA&A..47..481A}.

\section{Data}\label{ss_data}

\subsection{Galaxy Selection}
The galaxies in this study were selected from the Subaru/Hyper Suprime-Cam Subaru strategic program (HSC-SSP) survey datasets \citep{2018PASJ...70S...4A}.
The Subaru/HSC survey is a photometric survey with optical broad band filters $grizy$ and several narrow-band filters.
The survey comprises three layers, UltraDeep, Deep, and Wide, with different combinations of area and depth (see \citealt{2018PASJ...70S...4A} for details).
LBGs at $z\sim4-7$ are selected from the HSC datasets using the dropout selection technique \citep{2016ApJ...828...26M,2018PASJ...70S..35M,2018ApJS..237....5M,2019ApJ...883..183M,2018PASJ...70S..10O,2018PASJ...70S..11H,2018PASJ...70S..12T}, and some of the LBGs have been spectroscopically confirmed \citep{2016ApJ...828...26M,2018PASJ...70S..35M,2018ApJS..237....5M,2019ApJ...883..183M,2018PASJ...70S..10O}.
In this study, we focus on luminous $z\sim6$ LBGs near the end of the era of cosmic reionization, whose rest-frame UV spectra can be studied with optical spectrographs.

\begin{figure*}
\centering
\includegraphics[width=0.9\hsize]{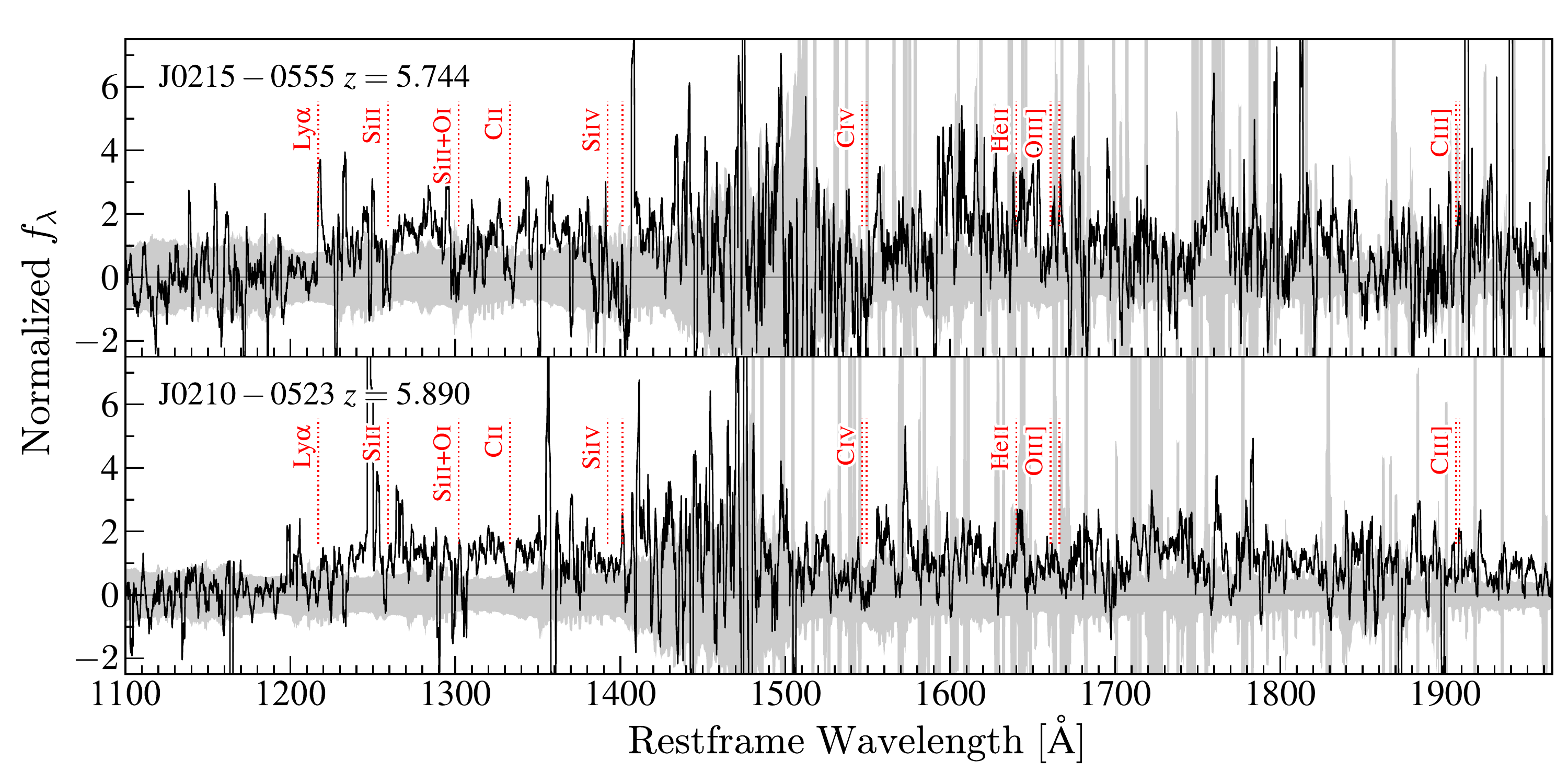}
\caption{Rest-frame VLT/X-shooter spectra of the targets J0215-0555  and J0210-0523.
The gray shaded regions indicate $1\sigma$ errors. Key diagnostic emission and absorption lines are indicated with red labels.}
\label{fig_X-shooter}
\end{figure*}

\subsection{Spectroscopic Observations}

\subsubsection{Subaru/FOCAS and GTC/OSIRIS}
A total of 31 luminous ($m_y\sim23-24\ \m{mag}$) LBGs at $z\sim6$ were spectroscopically confirmed with Subaru/FOCAS and GTC/OSIRIS.
\bluec{These 31 LBGs were classified as galaxies in \citet{2016ApJ...828...26M,2018PASJ...70S..35M,2018ApJS..237....5M,2019ApJ...883..183M} because the spectra do not show apparent AGN signatures such as broad or high ionization {\sc Nv} emission lines.}
The spectroscopic data were taken in the SHELLQs program \citep{2016ApJ...828...26M,2018PASJ...70S..35M,2018ApJS..237....5M,2019ApJ...883..183M}.
We will briefly explain the Subaru/FOCAS and GTC/OSIRIS observations and data reduction. 
Please see the SHELLQs papers \citep{2016ApJ...828...26M,2018PASJ...70S..35M,2018ApJS..237....5M,2019ApJ...883..183M} for more details.

The majority of spectra were taken with Subaru/FOCAS observations from 2015 November to 2019 May (IDs: S15B-070, S16B-071I, and S18B-011I; P.I.: Y. Matsuoka).
The observations were conducted with the VPH900 grism, SO58 order-sorting filter, and a $1.\carcsec0$ slit.
This configuration provides coverage from $\lambda_\m{obs}=0.75$ to $1.05\ \m{\mu m}$ with a spectral  resolution of $R\sim1200$\redc{, which covers Ly$\alpha$, Si{\sc ii}$\lambda$1260, O{\sc i}$\lambda$1302, Si{\sc ii}$\lambda$1304, C{\sc ii}$\lambda$1334, and Si{\sc iv}$\lambda\lambda$1394,1403 lines at $z\sim6$}.
All the observations were carried out on gray nights with a seeing of $0.\carcsec4-1.\carcsec0$.

Six galaxies were observed with GTC/OSIRIS from 2016 April to 2018 September (GTC19-15B, GTC4-16A, GTC8-17A, GTC3-18A, GTC8-18B, and GTC32-19A; PI: K. Iwasawa).
These data were taken with the R2500I grism and a $1.\carcsec0$ slit, providing coverage from  $\lambda_\m{obs}=0.74$ to $1.05\ \m{\mu m}$ with a spectral resolution of $R\sim1500$. The observations were carried out on both dark and gray nights with a seeing of $0.\carcsec6-1.\carcsec3$.

The FOCAS and OSIRIS exposures are summarized in Table \ref{tab_object}. All of the data was reduced using the Image Reduction and Analysis Facility
(IRAF). Bias correction, flat fielding with dome flats, sky subtraction, and 1d extraction were performed using standard techniques and wavelength calibration
was performed with reference to night sky emission lines. Flux calibration was based on white dwarf or B-type standard stars observed within a few days of the target observations. Slit losses were accounted for by scaling spectra to match the HSC magnitudes.

\subsubsection{VLT/X-shooter}
Our team was granted 4 half-nights in Visitor Mode in August-September 2017 initially to observe four LBGs among the HSC sample (J0215-0555, J0210-0523, J0219-0416 and J0210-0559, Program 099.A-0128, P.I.: R. Ellis). The planned observing time in each arm was chosen to maximize that in the NIR arm ($\m{UVB}=756\ \m{sec/exposure}$, $\m{VIS}=819\ \m{sec/exposure}$, and $\m{NIR}=900\ \m{sec/exposure}$). We adopted 1.0\arcsec$\times$11\arcsec, 0.9{\arcsec}$\times$11\arcsec, and 0.9{\arcsec}$\times$11\arcsec slits in the UVB, VIS, and NIR arms, respectively\redc{, covering Ly$\alpha$, Si{\sc ii}$\lambda$1260, O{\sc i}$\lambda$1302, Si{\sc ii}$\lambda$1304, C{\sc ii}$\lambda$1334, Si{\sc iv}$\lambda\lambda$1394,1403, He{\sc ii}$\lambda$1640, {\sc Oiii]}$\lambda\lambda$1661,1666, [{\sc Ciii}]$\lambda$1907, and {\sc Ciii}]$\lambda$1909 lines at $z\sim6$}. As a result of inclement weather, we concentrated our observations on the two brightest objects, namely J0215-0555\footnote{A fluorescent line {\sc Cii}*1335\AA\ has been detected in this spectra (see \citealt{2019MNRAS.487L..67B})}  ($m_z=23.8\ \m{mag}$) and  J0210-0523 ($m_z=24.0\ \m{mag}$) for which we secured 7.5 hours in good seeing  ($<$0.7'') for each target. Data was reduced using the ESOReflex pipeline \citep{2013A&A...559A..96F} version 3.2.0. The rest-frame spectra for these two sources are shown in Figure \ref{fig_X-shooter}. The systemic redshifts used to produce the rest-frame spectra were measured to be $z=5.744$ and $5.890$ for J0215-0555 and J0210-0523, respectively, following methods presented in Section \ref{ss_redshift}. \redc{The median continuum signal-to-noise ratio is $\sim0.2$/pixel.} As discussed below, the spectral resolution for these two sources ($R\sim8000$ in VIS and $\sim6000$ in NIR) complements that available for the larger sample of $31$ galaxies enabling us to test whether the lower resolution of the FOCAS and OSIRIS spectra is adequate for quantitative absorption line studies.
\bluec{The absence of high ionization lines such as {\sc Civ} indicate no strong AGN activity in J0215-0555 and J0210-0523.}

\begin{figure*}
\centering
\includegraphics[width=0.9\hsize]{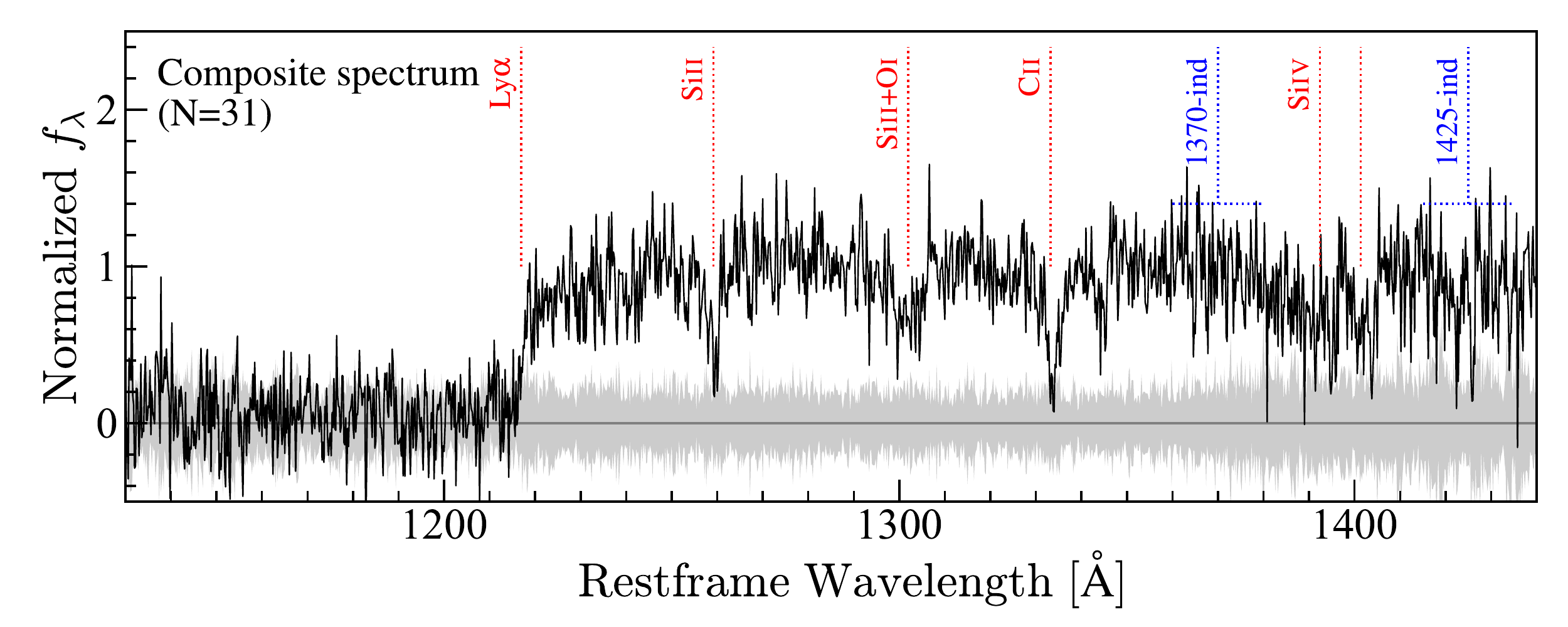}
\caption{Rest-frame composite spectrum derived from the 31 $z\sim6$ galaxies studied with FOCAS and OSIRIS.
The gray shaded region indicates $1\sigma$ uncertainties estimated by bootstrap resampling.
Key diagnostic interstellar and stellar absorption features are indicated with red and blue labels, respectively.
}
\label{fig_composite}
\end{figure*}

\section{Analysis}\label{ss_analysis}

\subsection{Redshift Determination}\label{ss_redshift}
In previous studies \citep{2016ApJ...828...26M,2018PASJ...70S..35M,2018ApJS..237....5M,2019ApJ...883..183M}, provisional redshifts for our 31 galaxies were determined using a variety of tracers including the Ly$\alpha$ emission line, interstellar absorption lines and the Lyman discontinuity. \bluec{It is well known that Ly$\alpha$ emission is offset in velocity from the systemic redshift of a galaxy \citep{2001ApJ...554..981P}. While various workers have attempted to infer the systemic redshift from Ly$\alpha$ using trends based on its line strength (e.g. \citealt{2003ApJ...584...45A}, \citealt{2018MNRAS.477.2098N}), these relations are poorly understood. Instead, for the purposes of stacking our spectra, we have determined their individual absorption line redshifts based on a cross-correlation technique \citep{1979AJ.....84.1511T}. This has the advantage of using all the absorption features in each of our spectra rather than simply those few recognised visually.
Each spectrum was zero-meaned and placed on a logarithmic wavelength scale and cross-correlated against the similarly-processed composite spectrum of 80 $z\sim4$ galaxies in \citet{2012ApJ...751...51J} smoothed to account for resolution differences}. \bluec{The absorption line redshift for each of our $z\sim6$ galaxies was determined using the peak in the cross-correlation function.} The same technique was used for the higher resolution X-shooter spectra of J0215-0555 and J0210-0523. \bluec{Since we will primarily be interested in the {\em depth} of absorption lines, small absorption line velocity offsets that might be present between the $z\sim4$ template and our $z\sim6$ spectra would lead to somewhat reduced covering fractions (discussed further in Section 4.2\redc{)}.}

The newly-derived redshifts are presented in Table \ref{tab_object} and typically accurate to $\Delta z<0.01$. Apart from the velocity offset of Ly$\alpha$, the results do not significantly differ from those reported in \citet{2016ApJ...828...26M,2018PASJ...70S..35M,2018ApJS..237....5M,2019ApJ...883..183M}.
In the specific case of J1211-0118 we adopt $z=6.0293\pm0.0002$ given this is more accurately determined from an ALMA observation \citep{2020ApJ...896...93H}. 

\subsection{Composite Spectrum}
A composite spectrum of the 31 galaxies studied by FOCAS and OSIRIS was generated by resampling the individual spectra to the rest-frame, normalizing their fluxes to unity, and median stacking. In this way, all individual spectra have equal weight irrespective of the target brightness or spectral signal/noise.
The error spectrum was estimated via bootstrap resampling. We created 100 alternate composites using the same reduced number of spectra drawn at random 
from the parent sample and adopted an error based upon the standard deviation of these composites. Both the composite and its error are presented in Figure \ref{fig_composite}.
\bluec{The median continuum signal-to-noise ratio is 5/pixel.}
The median redshift and UV luminosities of our sample of $z\sim6$ galaxies are $z=5.868$ and $M_\m{UV}=-22.9$ mag, respectively.
\citet{2019ApJ...883..183M} also constructed a composite spectrum of the same 31 galaxies using the provisional redshifts they measured. We note that the
equivalent widths (EW) of the absorption lines we analyze in our composite spectrum agree well with their measurements.

\begin{figure*}
\centering
\includegraphics[width=0.9\hsize]{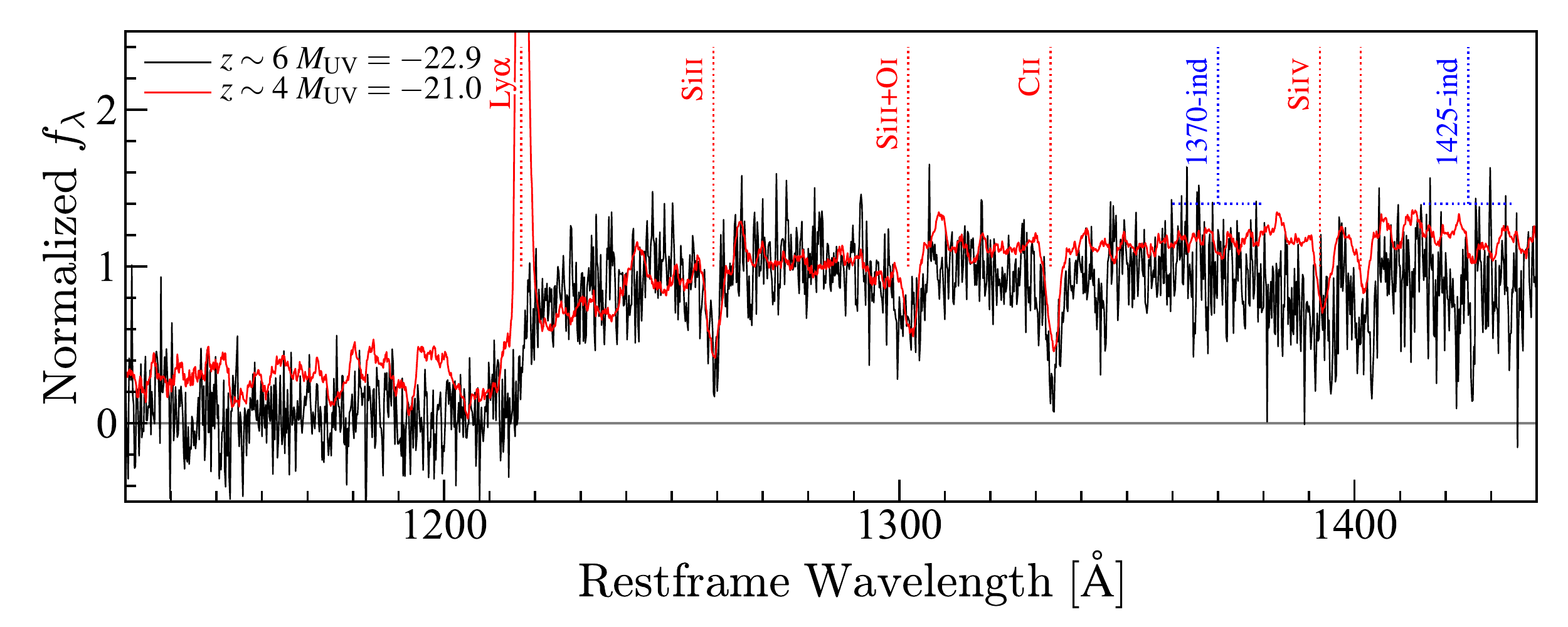}
\caption{Comparison of the spectrum in Figure \ref{fig_composite} (black) with the composite spectrum of 80 $z\sim4$ galaxies in \citet{2012ApJ...751...51J} (red).
The deeper absorption lines in our luminous $z\sim6$ galaxies indicates an increased low ionization gas covering fraction compared to the less luminous galaxies at $z\sim4$.}
\label{fig_composite_wJ12}
\end{figure*}

\section{Results}\label{ss_results}

\subsection{Absorption Features}
In both the composite spectrum (Figure \ref{fig_composite}) and individual spectra (Figure \ref{fig_X-shooter}), we can identify several interstellar absorption lines, including Si{\sc ii}$\lambda$1260, O{\sc i}$\lambda$1302, Si{\sc ii}$\lambda$1304, C{\sc ii}$\lambda$1334, and Si{\sc iv}$\lambda\lambda$1394,1403. 
In addition to these interstellar lines, we can also identify an absorption feature near $1425$ {\AA} consistent with a blend of stellar photospheric absorption lines arising from metal rich stars \citep{2004ApJ...615...98R}. In the individual X-shooter spectra of J0215-0555 and J0210-0523 that extend the wavelength coverage to longer wavelengths, we identify Si{\sc ii}$\lambda$1260, O{\sc i}$\lambda$1302, Si{\sc ii}$\lambda$1304, C{\sc ii}$\lambda$1334, Si{\sc iv}$\lambda\lambda$1394,1403, and C{\sc iv}$\lambda\lambda$1548,1550. However, even with the superior spectral resolution, the O{\sc i}$\lambda$1302 and Si{\sc ii}$\lambda$1204 absorption lines remain blended. Collectively, the presence of these absorption lines indicate that the interstellar media (ISM) of our 
luminous $z\sim6$ galaxies are metal enriched, which we discuss in more detail in Section \ref{ss_gasZ}.

To obtain quantitative measures of these absorption lines, we fit the individual and composite spectra with Gaussian profiles as follows:
\begin{equation}
I(\lambda)=I_0 \exp\left(-\frac{(\lambda-\lambda_0)^2}{2\sigma_\lambda^2}\right)+C,
\end{equation} 
where $I_0$, $\lambda_0$, $\sigma_\lambda$, and $C$ are maximum line strength, central wavelength, line dispersion, and continuum intensity, respectively.
\bluec{The {\sc Oi}$\lambda$1302 and Si{\sc ii}$\lambda$1304 lines are fitted only in the wavelength ranges corresponding to $v<200\ \m{km\ s^{-1}}$ and $v>-200\ \m{km\ s^{-1}}$, respectively, following \citet{2013ApJ...779...52J} \redc{in order to avoid contamination from the other feature}.}
Table \ref{tab_line} presents the resulting absorption line depths, line widths (FWHM) and rest-frame EW. Velocity offsets from the systemic redshift are typically $\sim200\ \m{km\ s^{-1}}$, consistent with the result derived from the $z\sim4$ composite spectrum \citep{2012ApJ...751...51J}. The absorption line widths are typically $\sim1000\ \m{km\ s^{-1}}$. 

In Figure \ref{fig_composite_wJ12} we compare our composite spectrum with that derived for the 80 $z\sim4$ galaxies in \citet{2012ApJ...751...51J}. 
\bluec{The absorption lines at $z\sim6$ are deeper than those at $z\sim4$, which we consider is more likely due to the increased luminosity of our sample ($M_\m{UV}\sim-23$ c.f. $M_\m{UV}\sim-21$ at $z\sim4$) rather than an evolutionary effect over a modest time interval. Several studies have reported larger EWs in more luminous galaxies \citep[e.g.,][]{2012ApJ...751...51J}. Specifically, over the redshift range $2<z<4$ \citet{2018ApJ...860...75D} report shallower absorption lines in higher redshift galaxies matched in the UV luminosity and stellar mass. This is opposite to the trend we would find if we interpreted our results as due to evolution.}

\subsection{Covering Fraction}\label{ss_fcov}
To estimate the neutral gas (H{\sc i}) covering fraction, we now measure the maximum absorption depth for LIS interstellar absorption lines following the methodology adopted by \cite{2013ApJ...779...52J}. Assuming the gas is distributed in a spherical shell, the absorption line profile is related to the covering fraction, $f_\m{c}$, via
\begin{equation}
\frac{I(v)}{C}=1-f_\m{c}(v)(1-\m{e}^{-\tau(v)}), \label{eq_fc_tau}
\end{equation}
where $C$ is the continuum intensity defined as the median flux over the wavelength range without absorption lines, and $\tau(v)$ is the optical depth of the relevant absorption line which is related to the column density as
\begin{equation}
\tau(v)=f \lambda \frac{\pi e^2}{m_\m{e} c}N(v)=f \lambda \frac{N(v)}{3.768\times10^{14}},
\end{equation}
where $f$, $\lambda$, and $N(v)$ are the ion oscillator strength, the transition wavelength in \AA, and the column density in $\m{cm}^2$, respectively.

In their study, \citet{2013ApJ...779...52J} used Si{\sc ii}$\lambda$$1260$, $1304$, and $1526$ and estimated $f_\m{c}$ as a function of velocity for several individual $z\sim4$ gravitationally-lensed galaxies. They found consistent results despite the different oscillator strengths indicating that the gas is optically thick. In the optically thick case ($\tau\gg1$), Equation (\ref{eq_fc_tau}) simplifies to
\begin{equation}\label{eq_fcov}
f_\m{c}(v)=1-\frac{I(v)}{C}.
\end{equation}
Note that the covering fraction derived in this way formally represents a lower limit.

\begin{figure}
\centering
\includegraphics[width=0.9\hsize]{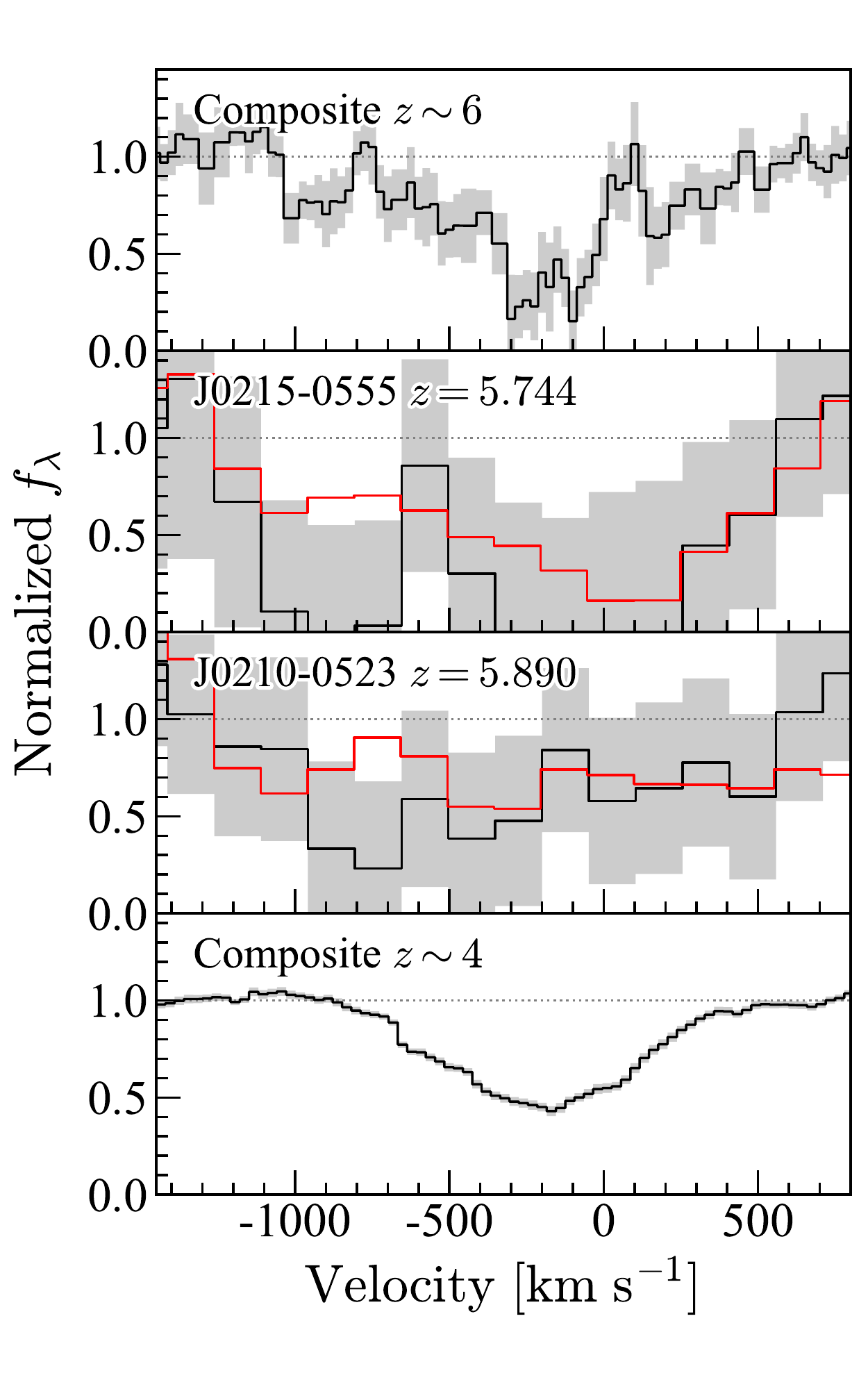}
\caption{The panels show averaged line profiles (black) with $1\sigma$ errors (shaded regions) for (from top to bottom) the $z\sim6$ composite spectrum, J0215-0555, J0210-0523, and the $z\sim4$ composite spectrum in \citet{2012ApJ...751...51J}. The black and red lines for the panels of J0215-0555 and J0210-0523 represent profiles derived from the X-shooter and FOCAS data, respectively. 
}
\label{fig_mean_abs}
\end{figure}

For the present analysis we constructed the average absorption line profile by calculating a weighted mean of LIS absorption lines as a 
function of velocity using only Si{\sc ii}$\lambda$1260 and C{\sc ii}$\lambda$1334. We chose not to use {O\sc i}$\lambda$1302 or Si{\sc ii}$\lambda$1304 transitions because they are blended. Figure \ref{fig_mean_abs} shows the averaged absorption line profiles for our $z\sim6$ composite spectrum, the two X-shooter individual spectra of J0215-0555 and J0210-0523, and the $z\sim4$ composite in \citet{2012ApJ...751...51J}. The estimated maximum absorption line depths are $0.85\pm0.16$, \redc{$1.00\pm0.32$}, $0.77\pm0.45$, and $0.57\pm0.03$ for our $z\sim6$ composite spectrum, J0215-0555, J0210-0523, and the $z\sim4$ composite spectrum, respectively, as summarized in Table \ref{tab_mes}. We note that if the systemic redshifts are not measured precisely, the absorption depth in our composite will be underestimated, indicating that true covering fraction could be larger. Finally we note that \cite{2013ApJ...779...52J} examined their average absorption line depths for a subsample based on the prominence of Ly$\alpha$ emission where they found a possible positive correlation. Out of our sample of 31 spectra, only 8 galaxies show prominent Ly$\alpha$ emission (with a median EW of 6.4 {\AA}). We found no significant difference between the absorption line statistics in this stack and that for the larger sample of 23 sources without Ly$\alpha$ emission.

To understand possible systematics in our estimates of the maximum absorption depth based on the lower-resolution FOCAS and OSIRIS spectra, we constructed  profiles for the specific cases of J0215-0555 and J0210-0523 and compared these with those based on the higher resolution X-shooter spectra (see red and black lines in Figure \ref{fig_mean_abs}). We found no significant difference indicating that the resolution of FOCAS ($R\sim1200$) is sufficient to estimate the maximum absorption line depth largely because of broad line widths involved (typically $\sim800-1000\ \m{km\ s^{-1}}$). We also degraded the X-shooter spectra to the lower resolution appropriate for the FOCAS/OSIRIS composite and similarly found no change in the absorption line depth. 

\begin{figure}
\centering
\includegraphics[width=0.99\hsize]{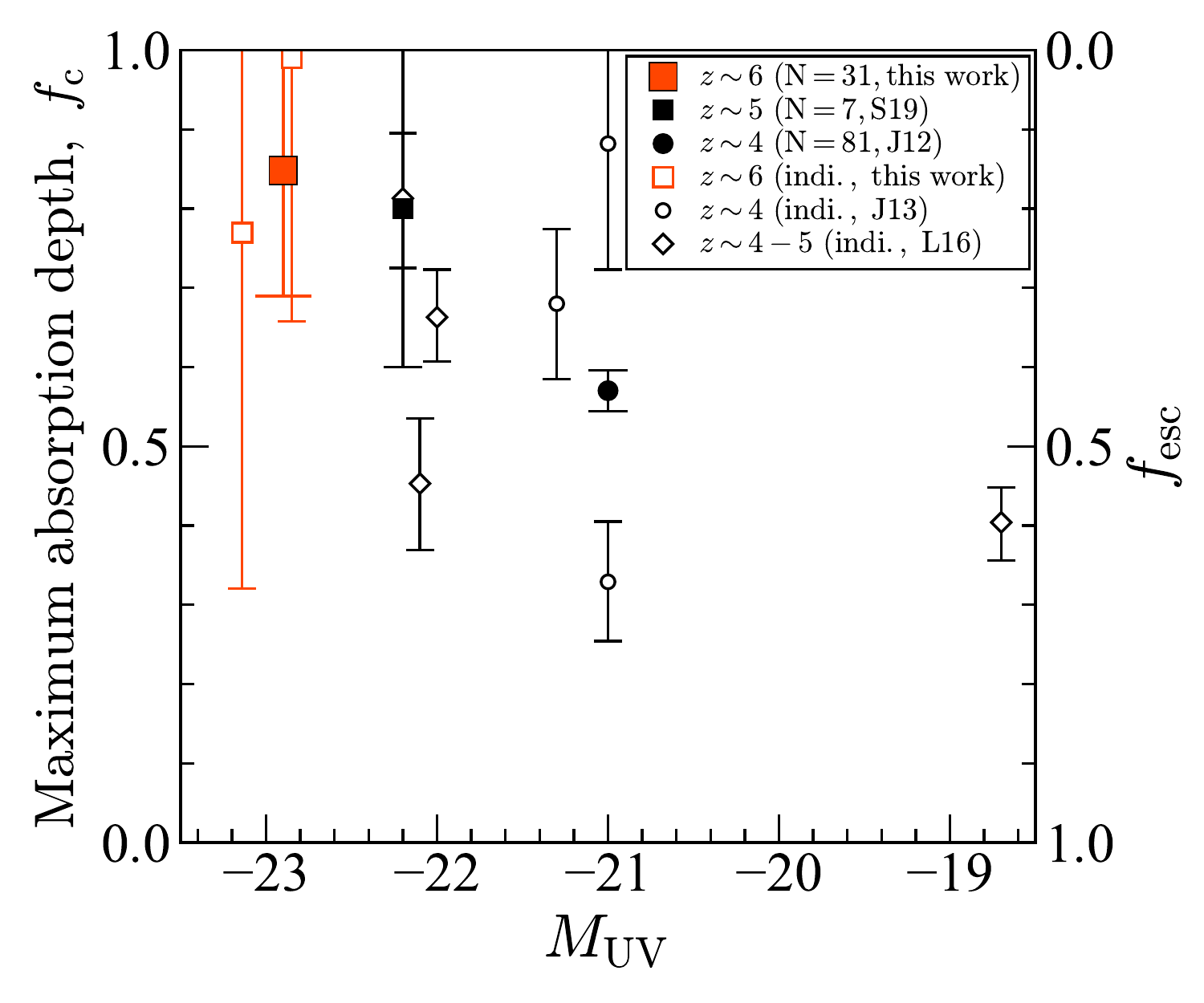}
\caption{Maximum absorption depth as a function of UV luminosity drawn from our $z\sim6$ composite spectrum (red filled square), J0215-0555 and J0210-0523 (open red square), composites at $z\sim5$ and $\sim4$ in \citet{2019ApJ...886...29S} and \citet{2012ApJ...751...51J}, respectively (black filled symbols), and individual galaxies in \citet{2013ApJ...779...52J} and \citet{2016ApJ...831..152L} (black open symbols).
As discussed in the text, these absorption line measurements correspond to a lower limit on $f_\m{c}$ and an upper limit on $f_\m{esc}$.}
\label{fig_abs_MUV}
\end{figure}

\begin{figure}
\centering
\includegraphics[width=0.99\hsize]{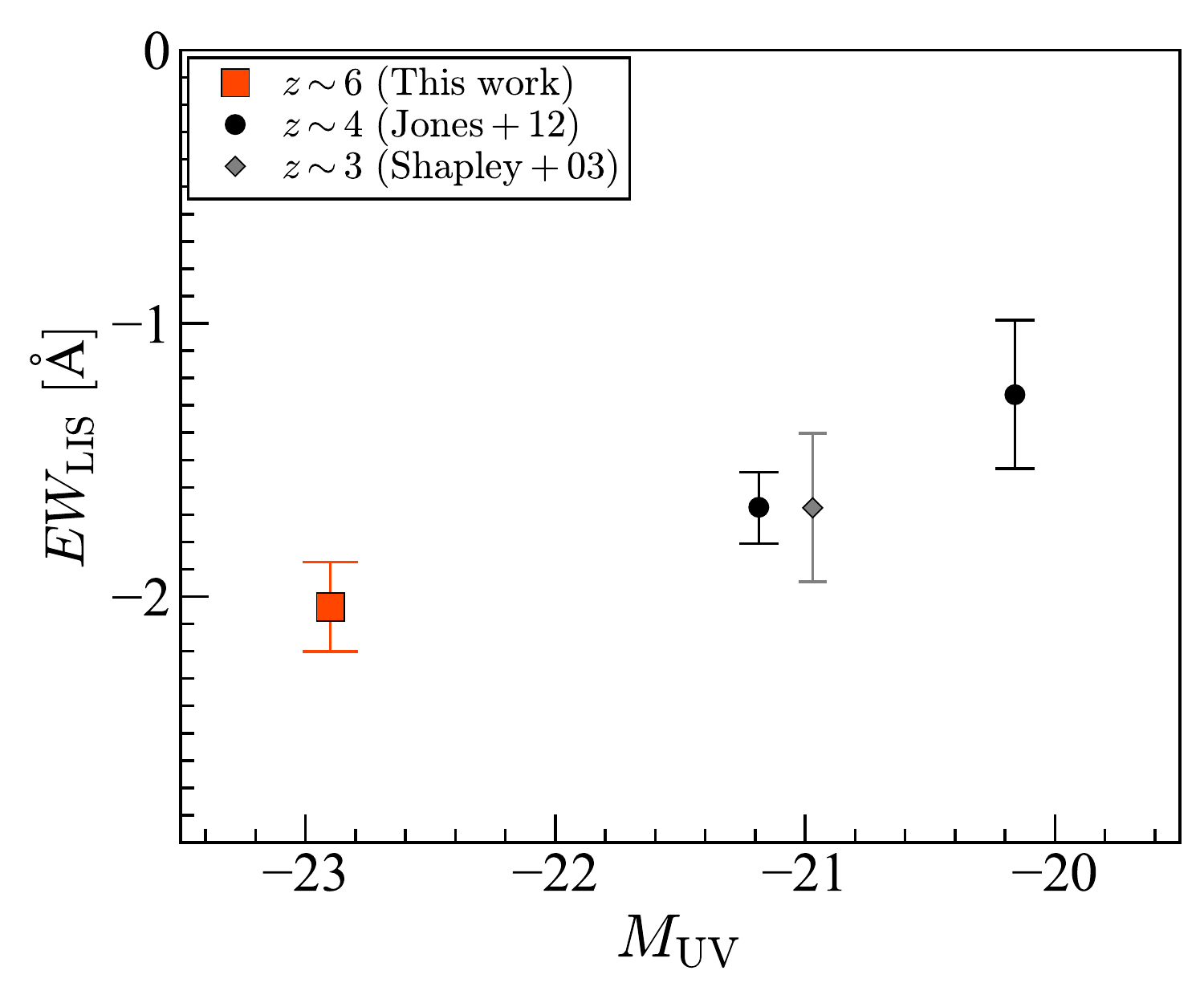}
\caption{\bluec{EW of LIS absorption, $EW_\m{LIS}$, as a function of UV luminosity drawn from our $z\sim6$ composite spectrum (red square), composites at $z\sim4$ and $\sim3$ in \citet{2012ApJ...751...51J} and \citet{2003ApJ...588...65S}, respectively (black and grey symbols).} \redc{$EW_\m{LIS}$ here is the averaged EW for Si{\sc ii}$\lambda$1260 and C{\sc ii}$\lambda$1334.}}
\label{fig_EWLIS_MUV}
\end{figure}

In Figure \ref{fig_abs_MUV}, we plot the estimated maximum absorption depth as a function of the UV luminosity. As discussed earlier, the maximum absorption 
depth of our $z\sim6$ galaxies is larger than those of less luminous sources at $z\sim4-5$ \citep{2012ApJ...751...51J,2013ApJ...779...52J,2016ApJ...831..152L,2019ApJ...886...29S}. This should not be interpreted as an evolutionary trend given the significantly different luminosities. The major implication is for a higher covering fraction of low ionization gas for our sample of $z\sim6$ luminous galaxies and thus a lower typical ionizing photon escape fraction.
\bluec{Figure \ref{fig_EWLIS_MUV} shows the EW of LIS absorption, $EW_\m{LIS}$, estimated from Si{\sc ii}$\lambda$1260 and C{\sc ii}$\lambda$1334. $EW_\m{LIS}$ \redc{is stronger} with increasing luminosity, consistent with the trend in the maximum absorption depth.}

\subsection{Gas-phase metallicity}\label{ss_gasZ}
Since the rest-frame optical emission lines normally used to estimate gas-phase metallicities are redshifted to the mid-infrared beyond reach of ground-based telescopes, we can only use empirically-calibrated relations between the oxygen abundance and interstellar absorption line measures \citep{2016ApJ...822...29F}.
For this measure, we use the O{\sc i}$\lambda$1302+Si{\sc ii}$\lambda$1304 complex (referred to as the Si{\sc iii}$\lambda$1300 complex in \citealt{2016ApJ...822...29F}), C{\sc ii}$\lambda$1334, Si{\sc iv}$\lambda\lambda$1394,1403, and C{\sc iv}$\lambda\lambda$1548,1550 (only for the X-shooter spectra).
We use the best-fit parameters in \citet{2016ApJ...822...29F}, but an uncertainty of assuming the best-fit parameters is much smaller than the statistical errors.
\bluec{We estimate the metallicity for each absorption line, and calculate the weighted mean as the fiducial value.}
In this way, we find gas-phase metallicities close to solar values.
Specifically, \redc{$12+\m{log(O/H)}=8.7^{+0.2}_{-0.2}$, $9.3^{+0.4}_{-0.5}$, and $8.6^{+0.5}_{-0.6}$ ($Z_\m{gas}=1.0^{+0.4}_{-0.3}$, $3.6^{+6.1}_{-2.5}$, and $0.8^{+1.5}_{-0.6}\ Z_\odot$)} for the $z\sim6$ composite spectrum, J0215-0555, and J0210-0523, respectively, as summarized in Table \ref{tab_mes}.

\begin{figure}
\begin{center}
\includegraphics[width=0.99\hsize]{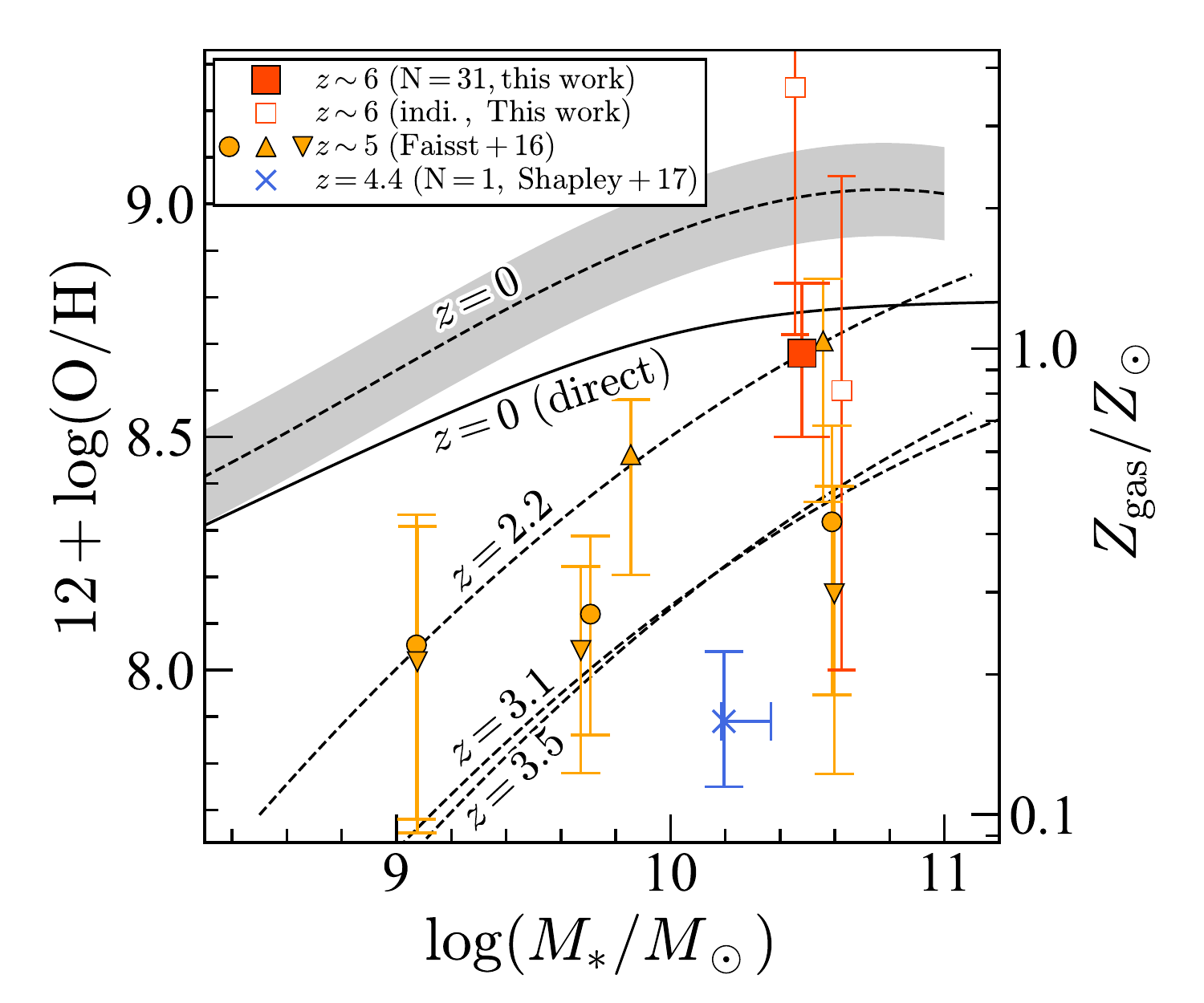}
\end{center}
\caption{Gas-phase metallicity as a function of the stellar mass based upon the $z\sim6$ composite spectrum and two individual spectra (J0215-0555 and J0210-0523, red filled and open squares), $z\sim5$ galaxies in \citet{2016ApJ...822...29F} (orange symbols where circles, upward triangles, and downward triangles refer to all galaxies, those with weak/no Ly$\alpha$, and with Ly$\alpha$, respectively). The blue cross represents a galaxy at $z=4.4$ in \citet{2017ApJ...846L..30S}. For comparison, \bluec{we plot relations at $z=0$ calibrated with the direct-temperature method \citep[][solid curve]{2020MNRAS.491..944C} and the strong line method \citep[][dashed curve with shades]{2008ApJ...681.1183K},} at $z=2.2$, $3.5$ \citep{2008A&A...488..463M}, and $z=3.1$ \citep{2009MNRAS.398.1915M}. \bluec{Stellar masses in \citet{2016ApJ...822...29F} are converted to the \citet{1955ApJ...121..161S} IMF.} }
\label{fig_Zgas_Ms}
\end{figure}

In Figure \ref{fig_Zgas_Ms}, we plot these gas-phase metallicities as a function of stellar mass where, in absence of Spitzer photometry, we derive approximate stellar masses based on an empirical relation with the UV luminosity given in \citet{2016ApJ...825....5S}. The resulting relation shows that 
our galaxies are already metal-enriched, and have metallicities comparable to those at the massive end of the $z\sim0$ mass-metallicity relation in \citet{2020MNRAS.491..944C}. Although these measures are comparable with those of $z\sim5$ galaxies with weak/no Ly$\alpha$ emission \citep{2016ApJ...822...29F}, they lie above mass metallicity relations at $z=3.1$ and $3.5$ in \citet{2008A&A...488..463M} and \citet{2009MNRAS.398.1915M} which seems surprising.
However, \citet{2014MNRAS.442..900N} point out that the published metallicities at $z=3.1-3.5$ would be higher if the assumed ionization parameter was closer to more recent estimates \citep[e.g.,][]{2013ApJ...769....3N,2014MNRAS.442..900N,2020ApJ...896...93H}. Future JWST rest-frame optical spectroscopy for a large sample of high redshift galaxies will resolve this issue and securely determine the gas-phase metallicities and the mass-metallicity relation of high redshift galaxies.

\begin{figure*}
\centering
\begin{minipage}{0.7\hsize}
\includegraphics[width=0.99\hsize]{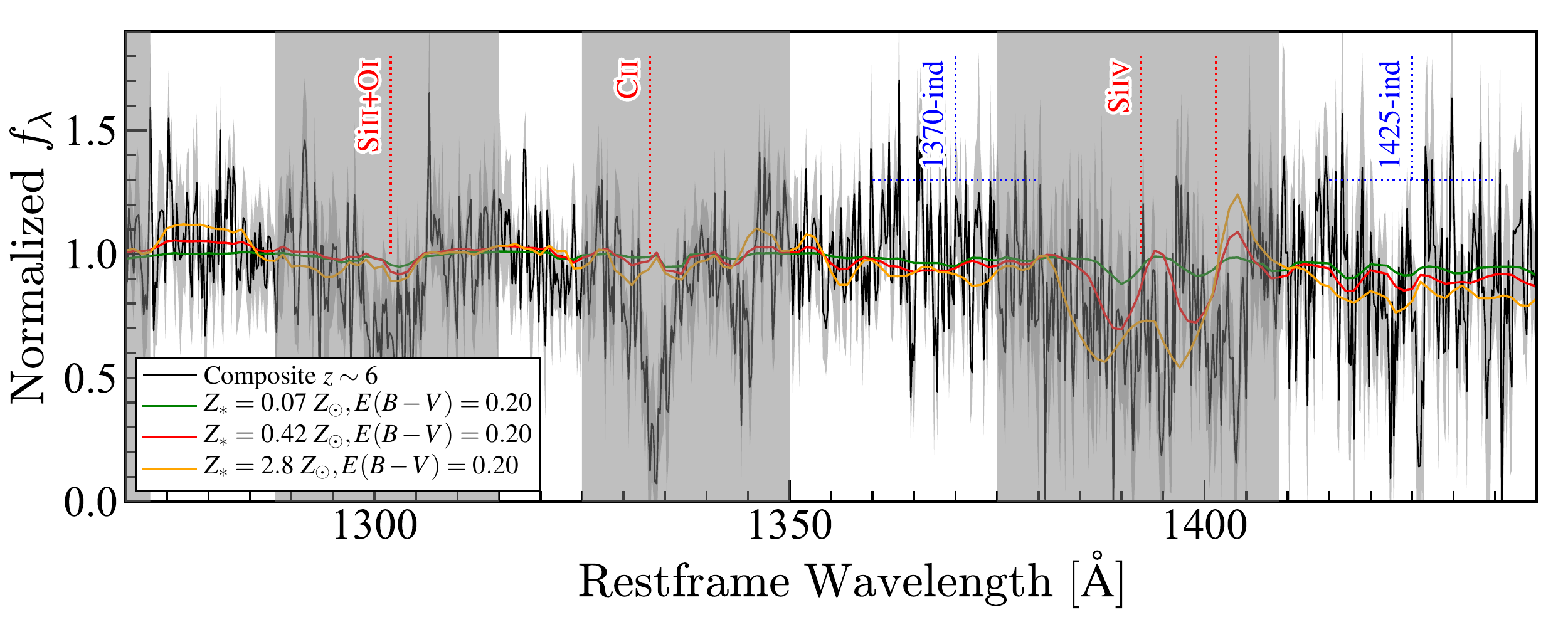}
\end{minipage}
\begin{minipage}{0.27\hsize}
\includegraphics[width=0.99\hsize]{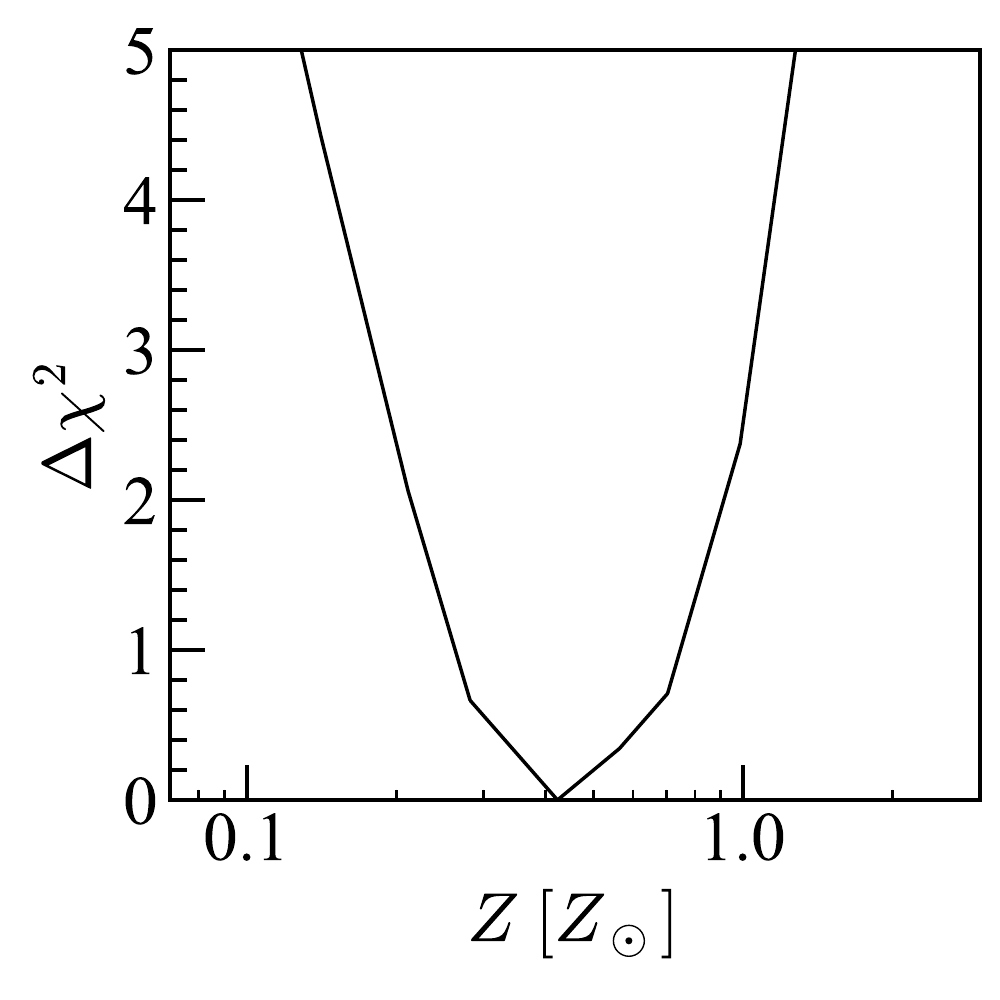}
\end{minipage}
\caption{Estimating the stellar metallicity at $z\sim6$. \bluec{The left panel compares the observed spectrum with the models.} The black line represents the $z\sim6$ composite spectrum and grey shaded regions indicate masked wavelength ranges not used  for fitting because of the prominent interstellar absorption lines. Green, red and orange curves represent BPASS model spectra (see text for details) with stellar metallicities of $Z_*=0.07$, $0.42$, and $3.0\ Z_\odot$, respectively, and a color excess of $E(B-V)=0.20$. Absorption features near 1370 and 1425 {\AA} become deeper with higher metallicities.
\bluec{The right panel shows the $\chi^2$ distribution for $E(B-V)=0.20$.} \redc{The y-axis is the difference from the best-fit $\chi^2$ value, $\Delta \chi^2=\chi^2 - \chi^2_\m{min}$.}
}
\label{fig_composite_wmodels}
\end{figure*}

\subsection{Stellar metallicity}
In our composite spectrum, we identify a significant feature near $1425\ ${\AA} which arises from photospheric absorption associated with stellar winds in massive stars. Its absorption line strength (and that of the blend near 1370 and 1425 {\AA}; the `1370' and `1425' indices) have been used as indicator of stellar metallicity \citep{2001ApJ...550..724L,2004ApJ...615...98R}.

\bluec{Although undoubtedly challenging to consider the possibility of deriving the stellar composition at $z\sim6$}, we compared our observed composite spectrum with model spectra assuming different metallicities following the techniques described by \citet{2016ApJ...826..159S} and \citet{2019MNRAS.487.2038C}. By simulating the entire spectrum, we can also use additional weaker features. For our model spectra we ran BPASSv2\footnote{https://bpass.auckland.ac.nz/} \citep{2016MNRAS.462.3302E,2016MNRAS.456..485S} sampling metalliticies over a grid of $Z_*=(0.001,0.002,0.003,0.004,0.006,0.008,0.010,0.014,\bluec{0.020,}$ $\bluec{0.030,0.040})$, adopting the \citet{1955ApJ...121..161S} IMF including binary stars and a continuous star formation history with a duration of $100\ \m{Myr}$.\footnote{Assuming a shorter duration of $10\ \m{Myr}$ does not change the conclusions, and assuming an older age does not affect our the results because the photospheric absorption line EWs near 1370 and 1425 {\AA} are saturated at $>30\ \m{Myr}$ \citep{2004ApJ...615...98R}}
As these models do not include the nebular continuum, we calculated the additional contribution using Cloudy \citep{1998PASP..110..761F,2017RMxAA..53..385F} version 17.01.
Assuming typical high redshift values for the electron density $n_\m{e}=300\ \m{cm^{-3}}$ and ionization parameter $\m{log}U=-2.8$ \citep{2016ApJ...826..159S},
this contributes an additional 10\% to the total far-UV continuum. The resulting model spectra were reddened following the \citet{2000ApJ...533..682C} dust extinction law parameterized by a color excess in the range of $0.00\leq E(B-V)\leq1.00$, and IGM attenuation applied following the approach given by \citet{2014MNRAS.442.1805I}. \bluec{Finally the model spectra are smoothed to various resolutions to match the velocity width.} Figure \ref{fig_composite_wmodels} shows a selection of models with $Z=0.07$, $0.42$, and $3.0\ Z_\odot$ with $E(B-V)=0.20$ alongside our composite spectrum.

To determine the best model, we calculated $\chi^2$ value for each model following
\begin{equation}
\chi^2=\sum_{i}\left(\frac{f_i^\m{model}-f_i^\m{obs}}{\sigma_i}\right)^2,
\end{equation}
where $f_i^\m{obs}$ and $\sigma_i$ are the observed flux and its error at each wavelength, and $f_i^\m{model}$ is the model spectrum normalized to the observed continuum level. The fit was confined to wavelength ranges of $1273<\lambda/\text{\AA}<1288$, $1315<\lambda/\text{\AA}<1325$, $1350<\lambda/\text{\AA}<1375$, and $1409<\lambda/\text{\AA}<1440$ which include absorption features near 1370 and 1425 {\AA} and avoid interstellar features. \bluec{Based on the $\chi^2$ values,} our best-fit stellar metallicity and color excess \bluec{with errors} are, respectively, $Z_*=0.42^{+0.28}_{-0.14}\ Z_\odot$ and $E(B-V)=0.20^{+0.06}_{-0.07}$ with a reduced $\chi^2$ value of $\chi^2/\nu=0.97$ (see Table \ref{tab_mes} and the red curve in Figure \ref{fig_composite_wmodels}).
\bluec{Naturally the error bar for our stellar metallicity estimate is larger than that in \citet{2019MNRAS.487.2038C} since our continuum signal-to-noise ratio is not as high as in their spectra.}

\begin{figure}
\begin{center}
\includegraphics[width=0.99\hsize]{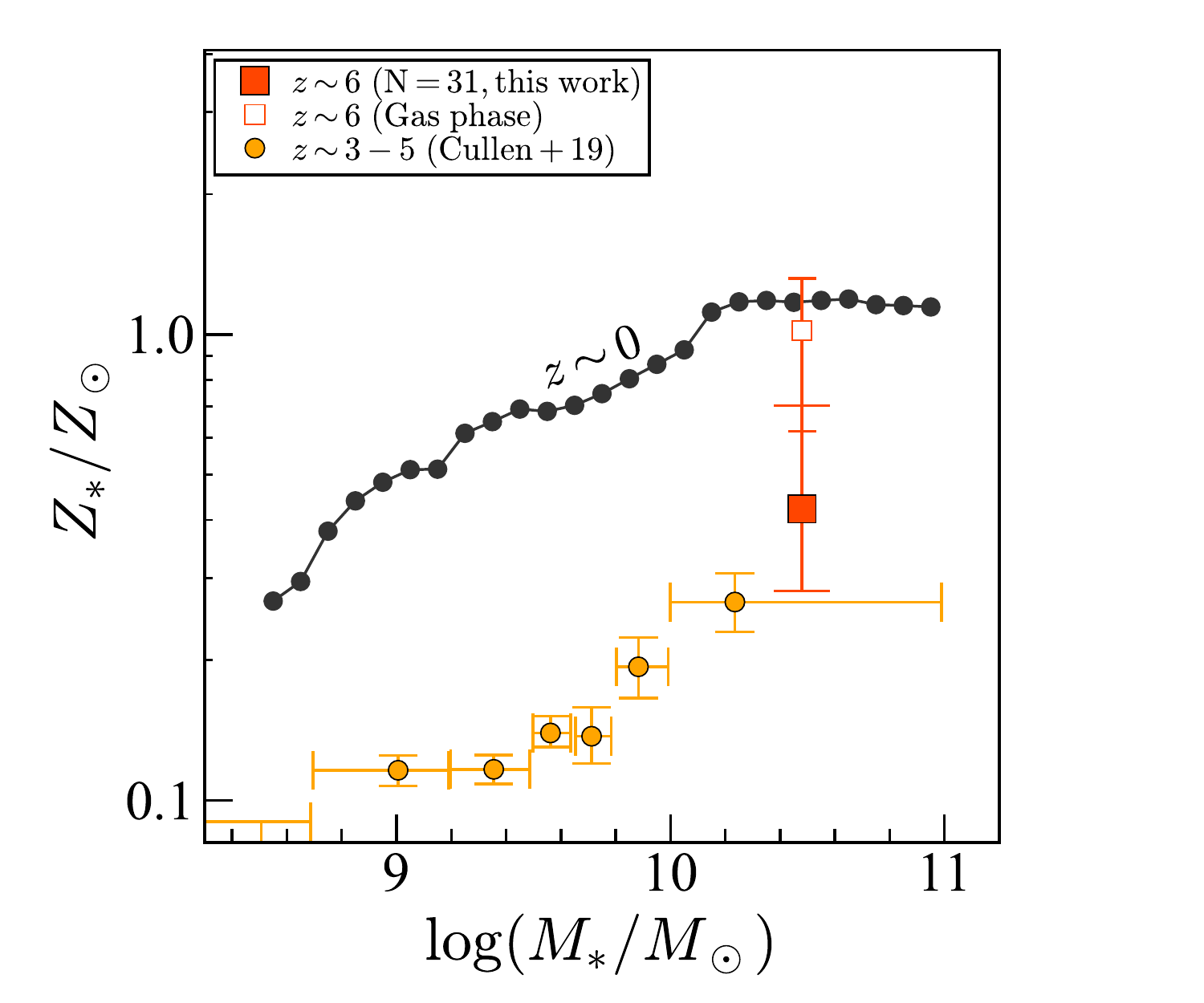}
\end{center}
\caption{Stellar metallicity as a function of the stellar mass based upon our $z\sim6$ galaxies (red filled square), $z\sim3-5$ galaxies from \citet{2019MNRAS.487.2038C} and $z\sim0$ from \citet{2017ApJ...847...18Z} (orange and black circles respectively).
The red open square represents the gas-phase metallicity based upon our $z\sim6$ galaxies (see Figure \ref{fig_Zgas_Ms}).}
\label{fig_Zs_Ms}
\end{figure}

In Figure \ref{fig_Zs_Ms}, we plot our estimated stellar metallicity as a function of the stellar mass. It can be seen that, for our $z\sim6$ luminous galaxies, our result agrees with an extrapolation of the mass-metallicity relation at $z\sim3-5$ from VANDELS in \citet{2019MNRAS.487.2038C} and is significantly lower than the $z\sim0$ relation \citep{2017ApJ...847...18Z}. Importantly, our stellar metallicity is lower than the gas-phase metallicity estimated in Section \ref{ss_gasZ}. We return to discuss this difference in Section \ref{ss_OFe}.

\subsection{Emission Lines}
\redc{The FOCAS and OSIRIS spectra cover Lya emission lines at $z\sim6$.
Among the 31 galaxies, only 8 galaxies show prominent Ly$\alpha$ emission with a median EW of 6.4 {\AA} \citep{2016ApJ...828...26M,2018PASJ...70S..35M,2018ApJS..237....5M,2019ApJ...883..183M}, indicating that Ly$\alpha$ emission is weak in our luminous galaxies.}
The X-shooter spectra cover the both optical and near-infrared wavelength region where, at high redshift, important rest-frame UV emission lines such as \bluec{Ly$\alpha$}, He{\sc ii}$\lambda$1640, {\sc Oiii]}$\lambda\lambda$1661,1666, [{\sc Ciii}]$\lambda$1907, and {\sc Ciii}]$\lambda$1909 might be visible. 
\bluec{We detect Ly$\alpha$ emission lines in J0215-0555 whose EW is $4.5\pm1.0$ \AA, consistent with \citet{2018PASJ...70S..35M} within $2\sigma$.}
We do not detect any of the \bluec{other} emission lines with typical \bluec{$2\sigma$ upper limits of the rest-frame EW of $<2$ \AA}. These non-detections are consistent with published trends for weaker emission in more luminous galaxies \citep[e.g.,][]{2003ApJ...588...65S,2018PASJ...70S..15S}. Since some of these lines are seen in metal-poor galaxies \citep[e.g.,][]{2018ApJ...859..164B,2018A&A...612A..94N}, this provides further evidence that our $z\sim6$ galaxies are metal-enriched.

\section{Discussion}\label{ss_discussion}

\subsection{Implication for Reionization}

We find that the average low ionization gas covering fractions for our luminous $z\sim6$ galaxies is significant, $f_\m{c} \ge 0.85\pm0.16$, and, as shown in Figure \ref{fig_abs_MUV}, \bluec{we consider this arises primarily due to a luminosity dependence}. Given the expected inverse relationship with respect to the ionizing photon escape fraction, i.e. $f_\m{esc} \simeq 1 - f_\m{c}$, and the fact the ionizing photon production efficiency for LBGs, $\xi_\m{ion}$, does not depend strongly on UV luminosity \citep[e.g.,][]{2016ApJ...831..176B}, our result implies \bluec{that the most luminous galaxies at $z\simeq6$ are not exceptional in their ionizing contribution, as we discuss further below}.

Recently, there has been much interest in the suggestion that galaxies of various luminosities made distinct contributions to the reionization process due, for example, to differing escape fractions or ionizing photon production rates. This contrasts with early articles (e.g. \citealt{2013ApJ...768...71R,2015ApJ...802L..19R}) which assumed a constant $f_\m{esc}$ and $\xi_\m{ion}$ independent of luminosity so that the process was governed by the most abundant, low-luminosity, sources. Even so, a high average fraction, $f_\m{esc}\simeq$ 0.2, was deemed necessary to complete reionization by $z\simeq6$. 

To resolve this possible `ionizing photon deficit', \citet{2019ApJ...879...36F} presented a model where the escape fraction is significantly higher in less luminous galaxies on the basis of simulations conducted by \citet{2015MNRAS.451.2544P}. In this model, $\sim80\%$ of the ionizing budget arises from sub-luminous galaxies with $M_\m{UV}>-16\ \m{mag}$. While possibly consistent with the present paper's results, Finkelstein's model predicts a relatively smooth evolution of the IGM neutral hydrogen fraction, e.g., $x_\m{HI}=0.3$ at $z=7.5$, which seems in tension with measures of QSO damping wings and Ly$\alpha$ fractions, $x_\m{HI}\simeq0.4-0.9$ \citep{2011MNRAS.416L..70B,2018Natur.553..473B,2018ApJ...856....2M,2019MNRAS.485.3947M,2019ApJ...878...12H}, as mentioned in \citet{2019ApJ...879...36F} and \citet{2020ApJ...892..109N}. 

The apparent rapid evolution of $x_\m{HI}$ over $6 < z < 7.5$ led 
\citet{2020ApJ...892..109N} to propose an alternative model where late reionization is governed by the most luminous galaxies with high star formation rate surface densities, $\Sigma_\m{SFR}$. Adopting a relationship whereby $f_\m{esc}\propto\Sigma_\m{SFR}^{0.4}$ \bluec{resulting in the higher escape fraction in more luminous galaxies}, $\gtrsim80\%$ of the ionizing budget is contributed by luminous galaxies with $M_\m{UV}<-18\ \m{mag}$, and the abrupt end of the reionization process as indicated by Gunn-Peterson troughs, QSO damping wings and Ly$\alpha$ fractions can be reconciled. \bluec{Although we emphasize it is not yet practical to conduct absorption line spectroscopy for a representative sample of the $z\simeq6$ galaxy population to $M_\m{UV}<-18\ \m{mag}$, the larger covering fraction we find for a more luminous subset does not provide any support for such a model.}

Ultimately, progress in addressing the relative roles of galaxies of different luminosities will depend on the accuracy of the various measures of the IGM neutral fraction, $x_\m{HI}$. \citet{2019ApJ...879...36F} discuss several reasons to be cautious when interpreting
QSO damping wing and Ly$\alpha$ fraction data, because the derived $x_\m{HI}$ measurements are very model dependent. If, as we suggest, escape fractions are indeed lower in luminous galaxies, the validity of current measures of $x_\m{HI}$ may need to be re-considered.
\redc{Other factors that may need to be reconsidered include the $\xi_\m{ion}$ dependence on UV luminosity and the AGN contribution.
Higher $\xi_\m{ion}$ in more luminous galaxies, or a significant contribution from AGN could reproduce the observed rapid evolution of $x_\m{HI}$, although current observations do not strongly support these.}

\subsection{Enhanced $\m{O/Fe}$ ratio}\label{ss_OFe}

We find that the gas-phase metallicity estimated from our $z\sim6$ composite spectrum, $Z_\m{gas}\simeq1.0\ Z_\odot$, is higher than the stellar metallicity, $Z_*\simeq0.4\ Z_\odot$. Such a difference is also seen in lower redshift galaxies \citep{2016ApJ...826..159S,2019MNRAS.487.2038C}. We note that the empirical relations used to estimate the gas-phase metallicity is calibrated with the oxygen abundance, $12+\m{log(O/H)}$, whereas the stellar metallicity is estimated from photospheric absorption in the rest-UV continuum dominated mainly by iron-peak element \citep{2016ApJ...826..159S}. Accordingly the difference between gas-phase and stellar metallicities indicates a non-solar elemental abundance pattern, i.e., a super-solar $\m{O/Fe}$ ratio.\footnote{It is safe to assume that gas-phase and stellar metallicities trace similar timescales because our stellar metallicity is estimated from the rest-UV spectrum dominated by short-lived massive stars.} From our two metallicity estimates in the $z\sim6$ composite spectrum, we derive an $\m{O/Fe}$ ratio of $\m{(O/Fe)}\simeq(2.4\bluec{\pm1.4})\times\m{(O/Fe)}_\odot$. \bluec{Although the error is large,} such a super-solar $\m{O/Fe}$ ratio has also been reported at lower redshifts; $\m{(O/Fe)}\sim3-7\times\m{(O/Fe)}_\odot$ at $z\sim2$ \citep{2016ApJ...826..159S,2018ApJ...868..117S,2019arXiv191210243T} and $\m{(O/Fe)}\gtrsim1.8\times\m{(O/Fe)}_\odot$ at $z\sim3-5$ \citep{2019MNRAS.487.2038C}.

A super-solar $\m{O/Fe}$ ratio is not surprising for galaxies whose ISM has been enriched primarily by Type II (core-collapse) supernovae (SNe) . Oxygen is a primary product of core-collapse SNe and therefore has a short formation timescale, while iron production occurs largely from Type Ia SNe which form $\sim1$ Gyr after star formation (Figure 1 in \citealt{2019A&ARv..27....3M}).
Quantitatively, the yield from core-collapse SNe is a super-solar $\m{O/Fe}$ ratio; $\m{(O/Fe)}\sim4-6\times\m{(O/Fe)}_\odot$ for the \citet{1955ApJ...121..161S} IMF \citep{2006NuPhA.777..424N}, $\sim3\times\m{(O/Fe)}_\odot$ for the \citet{2001MNRAS.322..231K} IMF \citep{2004ApJ...608..405C,2006ApJ...647..483L}. By contrast, the yield from Type Ia SNe is $\m{(O/Fe)}\sim0.03\times\m{(O/Fe)}_\odot$ \citep{1999ApJS..125..439I}. For our $z\sim6$ galaxies, seen $\sim1\ \m{Gyr}$ after the Big Bang, metal enrichment is therefore dominated by core-collapse SNe with yields of $\m{(O/Fe)}\sim3-6\times\m{(O/Fe)}_\odot$.

\subsection{Implication for star formation prior to $z\sim6$}\label{ss_W17}

The near-solar gas-phase metallicity we observe for our $z\sim6$ galaxies is higher than that predicted by recent theoretical simulations. For example, the IllustrisTNG simulation predicts a gas-phase metallicity of $Z_\m{gas}\simeq0.4-0.5\ Z_\odot$ for 
massive galaxies at $z=6$ with $\m{log}(M_*/M_\odot)=10.5$ \citep{2019MNRAS.484.5587T}. Although the FIRE simulation predicts the $z=6$ mass-metallicity relation only to masses of $\m{log}(M_*/M_\odot)=9.5$ \citep{2016MNRAS.456.2140M}, extrapolating the relation still implies $Z_\m{gas}\sim0.3\ Z_\odot$ at $\m{log}(M_*/M_\odot)=10.5$. As discussed in \citet{2019MNRAS.487.2038C}, these simulations suffer from uncertainties in the assumed stellar yields, the strength of galactic outflows and star formation histories. \bluec{Although there are several large uncertainties in the stellar mass and metallicity estimates,} the higher metallicities we observe may imply star formation began much earlier than predicted in the simulations.

To quantify the possibility of star formation prior to $z\sim6$, we compare our results with the analytic one-zone chemical evolution model presented in \citet[][hereafter W17]{2017ApJ...837..183W}. In this model, gas is enriched from both core-collapse and Type Ia SNe with yields given by \citet{2004ApJ...608..405C}, \citet{2006ApJ...647..483L}, and \citet{1999ApJS..125..439I}.
Type Ia SNe suffer a minimum delay time of $0.15\ \m{Gyr}$ and an $e$-folding timescale of $1.5\ \m{Gyr}$ that are fiducial values in W17. The star formation rate (SFR) is coupled to the gas mass ($M_\m{gas}$) with a gas depletion timescale, $t_\m{dep}=M_\m{gas}/SFR$, assumed to be constant. Here we adopt $t_\m{dep}=400\ \m{Myr}$ following the prescription of \citet{2017ApJ...837..150S} at $z=6$.
The gas outflow is assumed to be a constant multiple of the SFR with a mass-loading factor of $\eta=\dot{M}_\m{out}/SFR$, where we
assume $\eta=2$ (we discuss the sensitivity to this choice below). The case for three star formation histories are explicitly solved in W17, but here we consider the constant star formation history for illustrative purposes.

\begin{figure}
\begin{center}
\includegraphics[width=0.99\hsize]{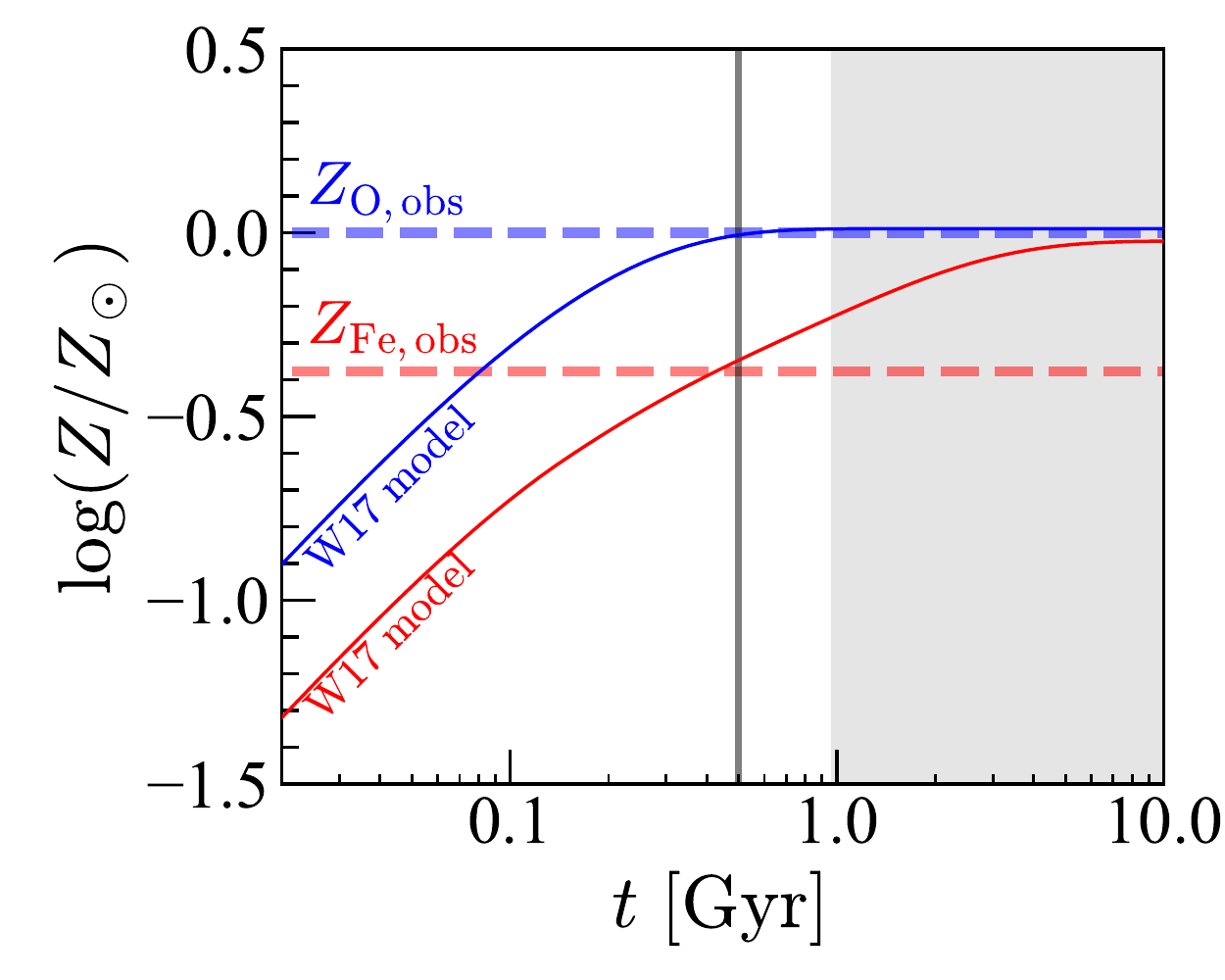}
\end{center}
\caption{The metallicity predicted in the chemical evolution model of \citet[][W17]{2017ApJ...837..183W} as a function of time after  initial star formation. Blue and red solid curves represent predicted oxygen and iron abundances assuming a constant star formation history with a gas depletion timescale of $t_\m{dep}=400\ \m{Myr}$ and the mass loading factor of $\eta=2$. Horizontal dashed lines indicate the observed gas-phase (oxygen) and stellar (iron) metallicities for our $z\sim6$ composite. Continuous star formation over $\sim500\ \m{Myr}$ is required to reproduce the observed metallicities. The gray-shaded region defines the parameter space rejected by the finite cosmic age at $z\sim6$ ($\sim1\ \m{Gyr}$).
}
\label{fig_W17}
\end{figure}

Figure \ref{fig_W17} shows the evolving abundances as a function of time from initial star formation. The oxygen abundance rapidly reaches a solar value in $\sim500\ \m{Myr}$ whereas for the iron abundance this takes $>1\ \m{Gyr}$. To reproduce the observed values in the case of a constant SFR we need a stellar age of $500\ \m{Myr}$. Decreasing $\eta$ pushes metal enrichment to earlier epochs, but a constant SFR is still needed over a duration of $\sim200\ \m{Myr}$ even for the case of zero outflow ($\eta=0$). While uncertainties remain, the metallicities we observe in our $z\sim6$ galaxies indicate enrichment that began up to $500\ \m{Myr}$ prior to $z\sim6$, corresponding to a formation redshift $z \sim10$, consistent with that implied by mature stellar populations inferred in higher redshift galaxies \citep{2018Natur.557..392H,2020arXiv200202968R}.

\section{Summary}\label{ss_conclusion}
We examine the absorption line spectra of a sample of 31 luminous Lyman break galaxies at redshift $z\sim6$ selected from the Subaru/Hyper Suprime-Cam Subaru strategic program for which spectra were taken using the Subaru/FOCAS and GTC/OSIRIS spectrographs. For two individual sources we present longer exposure data taken at higher spectral resolution with VLT/X-shooter.
Using these data, we demonstrate the practicality of stacking our lower resolution data to measure the depth of absorption lines and thereby to probe both the covering fraction of low ionization gas and the mean gas-phase and stellar metallicity near the end of cosmic reionization.
Our major findings are summarized below.

\begin{enumerate}

\item
We identify interstellar absorption lines of Si{\sc ii}, O{\sc i}, C{\sc ii}, and Si{\sc iv} in the composite and individual spectra of our $z\sim6$ galaxies. These absorption lines are deep and broad, typically with EWs of $\sim2-3\ ${\AA} and line widths of $\sim1000\ \m{km\ s^{-1}}$.
As a result of these broad absorption lines, we verify using our higher resolution X-shooter spectra that the spectral resolution of the FOCAS and OSIRIS data is sufficient to reliably estimate the depth of absorption.

\item
We find a maximum absorption depth of $0.85\pm0.16$ in the composite spectrum of our luminous ($M_\m{UV}\sim-23$ mag) galaxies. This is deeper than that of less luminous ($M_\m{UV}\sim-21$ mag) galaxies at $z\sim4$ and indicates a higher gas covering fraction and, by implication, \redc{a lower ionizing photon escape fraction. Our result suggests that the most luminous galaxies may not play a prominent role in concluding reionization.} Our result tends to support earlier models where the bulk of the ionizing photons arise from lower luminosity galaxies (e.g. \citealt{2013ApJ...768...71R,2015ApJ...802L..19R}, \citealt{2019ApJ...879...36F}). Since the predictions from these models in tension with model-dependent measurements of the neutral fraction based on the damping wing in QSO spectra and the incidence of Ly$\alpha$ emission in color-selected galaxies, it is important to understand the accuracy of such measures, \redc{as well as other factors such as the $\xi_\m{ion}$ dependence on UV luminosity and the AGN contribution}.

\item
We estimate the gas-phase metallicity of our galaxies using empirical relations linking these to the EWs of interstellar absorption lines. We find abundances close to solar, indicating that our luminous galaxies are already metal enriched at $z\sim6$. Similarly, we estimate a lower stellar metallicity of  $\sim0.4$ solar, which we interpret in terms of a super-solar O/Fe ratio given the stellar and gas-phase metallicities trace different elements enriched by core-collapse SNe. A comparison with the one-zone chemical evolution model in \citet{2017ApJ...837..183W} indicates that metal enrichment in our $z\sim6$ galaxies began up to $500\ \m{Myr}$ prior to $z\sim6$, corresponding to a formation redshift $z\sim10$.

\end{enumerate}

Our study provides a valuable impetus for more detailed absorption line studies of individual $z>6$ galaxies which will become practical with the James Webb Space Telescope.

\acknowledgments
We thank the anonymous referee for a careful reading and valuable comments that improved the clarity of the paper.
We thank Tucker Jones for sending us his composite $z\sim4$ spectrum, Andreas Faisst for providing parameter sets in his Monte Carlo simulations, and acknowledge useful discussions with Aayush Saxena, John Chisholm, \redc{Masami Ouchi, and Akio Inoue}.
YH acknowledges support from the JSPS KAKENHI grant No. 19J01222 through the JSPS Research Fellowship for Young Scientists.
RSE acknowledges funding from the European Research Council (ERC) under the European Union's Horizon 2020 research and innovation programme (grant agreement No 669253). 
NL acknowledges financial support from the Kavli Foundation. 

\begin{deluxetable*}{ccccccc}
\tablecaption{Summary of Objects Used}
\tablehead{\colhead{Name} & \colhead{$z$} & \colhead{$M_\mathrm{UV}$} & \colhead{$\mathrm{log}(M_*/M_\odot)$}  & \colhead{Instrument} & \colhead{$T_\mathrm{exp}$ (mins)} & \colhead{Ref.} \\
\colhead{(1)}& \colhead{(2)}& \colhead{(3)}& \colhead{(4)} &  \colhead{(5)} & \colhead{(6)} & \colhead{(7)}}
\startdata
J0215$-$0555 & $5.744\pm0.006$ & $-22.85$ & $10.5$ & X-shooter & 450 & This work\\
&  &  & & FOCAS & 220 & M16\\
J0210$-$0523 & $5.890\pm0.006$ & $-23.14$ & $10.6$ & X-shooter & 450 & This work\\
  &   &  &   & FOCAS   & 83  & M16\\
J0219$-$0416 & $5.973\pm0.006$ & $-22.56$ & $10.3$ & FOCAS & 80 & M16\\
J0210$-$0523 & $5.890\pm0.006$ & $-23.14$ & $10.6$ & FOCAS & 83 & M16\\
J0857$+$0142 & $5.827\pm0.006$ & $-22.71$ & $10.4$ & FOCAS & 100 & M16\\
J0848$+$0045 & $5.781\pm0.004$ & $-23.04$ & $10.5$ & FOCAS & 140 & M16\\
J1628$+$4312 & $6.020\pm0.008$ & $-22.90$ & $10.5$ & FOCAS & 170 & M18a\\
J1211$-$0118 & $6.0293\pm0.0002^*$ & $-23.23$ & $10.6$ & OSIRIS & 60 & M18a\\
J1630$+$4315 & $6.027\pm0.008$ & $-22.95$ & $10.5$ & FOCAS & 45 & M18a\\
J2233$+$0124 & $6.004\pm0.008$ & $-22.52$ & $10.3$ & FOCAS & 60 & M18a\\
J0212$-$0158 & $6.012\pm0.006$ & $-23.72$ & $10.9$ & OSIRIS & 60 & M18a\\
J0218$-$0220 & $5.867\pm0.015$ & $-22.94$ & $10.5$ & FOCAS & 60 & M18a\\
J0159$-$0359 & $5.781\pm0.004$ & $-22.78$ & $10.4$ & FOCAS & 60 & M18a\\
J2237$-$0006 & $5.777\pm0.005$ & $-22.37$ & $10.2$ & FOCAS & 100 & M18a\\
J1428$+$0159 & $6.006\pm0.004$ & $-24.30$ & $11.2$ & OSIRIS & 30 & M18b\\
J0917$-$0056 & $6.006\pm0.004$ & $-23.60$ & $10.8$ & OSIRIS & 120 & M18b\\
J0212$-$0315 & $5.916\pm0.005$ & $-22.85$ & $10.5$ & FOCAS & 50 & M18b\\
J0212$-$0532 & $5.901\pm0.005$ & $-22.42$ & $10.2$ & FOCAS & 50 & M18b\\
J2311$-$0050 & $5.904\pm0.008$ & $-22.72$ & $10.4$ & FOCAS & 50 & M18b\\
J1609$+$5515 & $5.868\pm0.008$ & $-22.41$ & $10.2$ & FOCAS & 80 & M18b\\
J1006$+$0300 & $5.863\pm0.015$ & $-22.98$ & $10.5$ & FOCAS & 45 & M18b\\
J0914$+$0442 & $5.851\pm0.006$ & $-23.79$ & $10.9$ & FOCAS & 50 & M18b\\
J0219$-$0132 & $5.780\pm0.003$ & $-22.25$ & $10.2$ & FOCAS & 30 & M18b\\
J0915$-$0051 & $5.664\pm0.006$ & $-22.60$ & $10.3$ & FOCAS & 75 & M18b\\
J135348.55$-$001026.5 & $6.176\pm0.004$ & $-24.76$ & $11.4$ & OSIRIS & 60 & M19\\
J144216.08$+$423632.5 & $6.016\pm0.015$ & $-22.93$ & $10.5$ & FOCAS & 30 & M19\\
J092117.65$+$030521.5 & $5.970\pm0.011$ & $-22.76$ & $10.4$ & FOCAS & 30 & M19\\
J115755.51$-$001356.2 & $5.867\pm0.007$ & $-22.98$ & $10.5$ & FOCAS & 40 & M19\\
J123841.97$-$011738.8 & $5.782\pm0.004$ & $-23.37$ & $10.7$ & FOCAS & 30 & M19\\
J162657.22$+$431133.0 & $5.819\pm0.003$ & $-22.78$ & $10.4$ & FOCAS & 50 & M19\\
J020649.98$-$020618.2 & $5.721\pm0.013$ & $-23.83$ & $10.9$ & OSIRIS & 60 & M19

\enddata
\tablecomments{(1) Object Name.
(2) Redshift.
(3) Absolute UV magnitude.
(4) \bluec{Stellar mass estimated from the UV magnitude with the relation at $z\sim6$ in \citet{2016ApJ...825....5S}.}
(5) Instrument.
(6) Exposure time.
(7) Reference (M16: \citealt{2016ApJ...828...26M}, 
M18a: \citealt{2018PASJ...70S..35M},
M18b: \citealt{2018ApJS..237....5M},
M19: \citealt{2019ApJ...883..183M}).\\
$^*$The systemic redshift of J1211$-$0118 is derived from ALMA observations in \citet{2020ApJ...896...93H}. 
}
\label{tab_object}
\end{deluxetable*}

\begin{deluxetable*}{cccccc}
\tablecaption{Properties of Absorption and Emission Lines}
\tablehead{\colhead{Name} & \colhead{Ion} & \colhead{$\lambda_\mathrm{rest}$}  & \colhead{$I_0/C$} & \colhead{FWHM} & \colhead{$EW_0$} \\
 & & \colhead{(\AA)}  &  & \colhead{$\mathrm{(km\ s^{-1}}$)} & \colhead{(\AA)} \\
\colhead{(1)} & \colhead{(2)}& \colhead{(3)}& \colhead{(4)} &  \colhead{(5)} & \colhead{(6)}}
\startdata
Composite spectrum & Ly$\alpha$   & 1215.69 & \nodata & \nodata &  $<1.4$ \\
& Si {\sc ii}  & 1260.42 & $-0.6 \pm 0.1$ &  $761 \pm 189$ &  $-2.1 \pm 0.5$ \\
& O {\sc i}   & 1302.17 & $-0.4 \pm 0.3$ &  $1483 \pm 950$ &  $-2.7 \pm 0.6$ \\
& Si {\sc ii}  & 1304.37 & $-0.4 \pm 0.1$ &  $761 \pm 189$ &  $-1.4 \pm 0.4$ \\
& C {\sc ii}   & 1334.53 & $-0.8 \pm 0.2$ &  $628 \pm 204$ &  $-2.4 \pm 0.5$ \\
& Si {\sc iv} & 1393.76 & $-0.5 \pm 0.1$ &  $1189 \pm 232$ &  $-3.1 \pm 0.7$ \\
& Si {\sc iv} & 1402.77 & $-0.4 \pm 0.1$ &  $1189 \pm 232$ &  $-2.5 \pm 0.5$ \\
J0215$-$0555 & Ly$\alpha$   & 1215.69 & \nodata &  $212 \pm 112$ &  $4.5 \pm 1.0$ \\
& Si {\sc ii}  & 1260.42 & $-1.1 \pm 0.6$ &  $1310 \pm 662$ &  $-6.0 \pm 2.0$ \\
& O {\sc i}   & 1302.17 & $-0.9 \pm 3.3$ &  $420 \pm 339$ &  $-1.8 \pm 1.1^*$ \\
& Si {\sc ii}  & 1304.37 & $-1.1 \pm 2.3$ &  $1310 \pm 662$ &  $-3.9 \pm 2.2^*$ \\
& C {\sc ii}   & 1334.53 & $-1.1 \pm 1.2$ &  $1000 \pm 577$ &  $-4.5 \pm 1.7$ \\
& Si {\sc iv} & 1393.76 & $-1.1 \pm 3.0$ &  $998 \pm 213$ &  $-4.7 \pm 2.0$ \\
& Si {\sc iv} & 1402.77 & $-1.1 \pm 1.8$ &  $998 \pm 213$ &  $-5.2 \pm 1.1$ \\
& C {\sc iv} & 1548.20 & $-0.9 \pm 0.5$ &  $1192 \pm 918$ &  $-5.1 \pm 1.0$ \\
& C {\sc iv} & 1550.78 & $-0.9 \pm 0.6$ &  $1192 \pm 918$ &  $-5.6 \pm 1.9$ \\
                        & He {\sc ii} & 1640.42 & \nodata & \nodata  & $<1.5$\\
                        & O {\sc iii}] & 1660.81 & \nodata & \nodata  & $<1.7$\\
                        & O {\sc iii}] & 1666.15 & \nodata & \nodata  & $<1.8$\\
                        & [C {\sc iii}] & 1906.68 & \nodata & \nodata  & $<54.1^{\dagger}$\\
                        & C {\sc iii}] & 1908.73 & \nodata & \nodata  & $<3.1$\\
J0210$-$0523 & Ly$\alpha$   & 1215.69 & \nodata & \nodata &  $<10.8$ \\
& Si {\sc ii}  & 1260.42 & \nodata & \nodata  &  $<4.2$ \\
& O {\sc i}   & 1302.17 & $-1.1 \pm 3.1$ &  $560 \pm 488$ &  $-2.9 \pm 0.9$ \\
& Si {\sc ii}  & 1304.37 & $-0.9 \pm 1.9$ &  $834 \pm 233$ &  $-3.4 \pm 0.7$ \\
& C {\sc ii}   & 1334.53 & $-0.9 \pm 0.2$ &  $948 \pm 452$ &  $-3.8 \pm 1.6$ \\
& Si {\sc iv} & 1393.76 & \nodata & \nodata  &  $<3.0$ \\
& Si {\sc iv} & 1402.77 & \nodata & \nodata  &  $<1.8$ \\
& C {\sc iv} & 1548.20 & $-0.8 \pm 0.5$ &  $815 \pm 730$ &  $-3.4 \pm 1.0$ \\
& C {\sc iv} & 1550.78 & $-0.9 \pm 0.4$ &  $815 \pm 730$ &  $-3.0 \pm 1.0$ \\
                        & He {\sc ii} & 1640.42 & \nodata & \nodata  & $<3.1$\\
                        & O {\sc iii}] & 1660.81 & \nodata & \nodata  & $<2.6$\\
                        & O {\sc iii}] & 1666.15 & \nodata & \nodata  & $<1.9$\\
                        & [C {\sc iii}] & 1906.68 & \nodata & \nodata  & $<2.5$\\
                        & C {\sc iii}] & 1908.73 & \nodata & \nodata  & $<3.1$
\enddata
\tablecomments{(1) Object Name.
(2) Ion.
(3) Rest-frame vacuum wavelength.
(4) Amplitude of the line.
(5) Width of the line corrected for the instrumental broadening.
(6) Equivalent width of the line. The upper limit is \bluec{$2\sigma$}.}
\tablenotetext{$*$}{\bluec{Although each absorption line is not detected beyond $2\sigma$, we show the best estimates because the combined signal-to-noise ratio is $2.3$.}}
\tablenotetext{$\dagger$}{The upper limit of [C {\sc iii}]1907 in J0215$-$0555 is relatively weak because the wavelength of the line is contaminated by a night sky OH line.}
\label{tab_line}
\end{deluxetable*}

\begin{deluxetable*}{ccccccc}
\tablecaption{Measurements for $z\sim6$ Galaxies}
\tablehead{
\colhead{Name} & \colhead{$z$} & \colhead{$M_\m{UV}$} & \colhead{$\m{log}(M_\mathrm{*}/M_\odot)$}  & \colhead{Max. abs. depth}  & \colhead{$12+\m{log(O/H)}$}  & \colhead{$Z_*/Z_\odot$} \\
\colhead{(1)} & \colhead{(2)}& \colhead{(3)}& \colhead{(4)} & \colhead{(5)} &  \colhead{(6)} &  \colhead{(7)}}
\startdata
$z\sim6$ composite & $5.868$ & $-22.90$ & $10.5$ & $0.85\pm0.16$ & $8.68^{+0.15}_{-0.18}$ & $0.42^{+0.28}_{-0.14}$\\
J0215$-$0555 & $5.744$ & $-22.85$ & $10.5$ & \redc{$1.00\pm0.32$} & $9.25^{+0.43}_{-0.53}$ & \nodata\\
J0210$-$0523 & $5.890$ & $-23.14$ & $10.6$ & $0.77\pm0.45$ & $8.60^{+0.46}_{-0.60}$ & \nodata
\enddata
\tablecomments{(1) Object Name.
(2) Redshift.
(3) Absolute UV magnitude.
(4) Stellar mass estimated from the UV magnitude with the empirical relation in \citet{2016ApJ...825....5S}.
(5) Maximum absorption depth.
(6) Gas-phase metallicity.
(7) Stellar metallicity}
\label{tab_mes}
\end{deluxetable*}

\bibliographystyle{apj}
\bibliography{apj-jour,reference}

\end{document}